\documentclass[prl,amsmath,amssymb,showpacs,preprint]{revtex4}

\usepackage{amsmath}
\usepackage{mathrsfs}
\usepackage{psfrag}
\usepackage{pstricks}
\usepackage{epsfig}  
\usepackage{amsmath}
\usepackage{array}
\usepackage{float}
\usepackage{latexsym}
\usepackage{mathrsfs}
\usepackage{pdflscape}
\usepackage{pstricks}
\usepackage{pst-node}
\usepackage{pst-plot}
\usepackage{subfigure}
\usepackage{tensor}
\usepackage{textcomp}
\usepackage{verbatim}
\usepackage{marvosym}

\newcommand{\bra}[1]{\langle #1 \vert}
\newcommand{\ket}[1]{\vert #1 \rangle}
\newcommand{\vv}{\mathrm{v}}

\def \kostelecky{Kosteleck\'{y}}

\begin{document}

\title{The Standard Model Extension For a Modified Ives-Stillwell Test}
\author{J. P. Cotter} 
\author{B. T. H. Varcoe} 
\email{b.varcoe@quantuminfo.org}
\affiliation{ Department of Physics and Astronomy,\\ The University of Leeds, \\ W. Yorkshire, LS2 9JT, UK} 

\begin{abstract}
In this paper we present the full theoretical model of a modified Ives-Stillwell experiment where counter propagating lasers are used to form a narrow interference fringe when the lasers form a double resnonance. This narrow resoannce can be as small as 1 Hz wide and  therefore provides a connection between the atomic resonance in its rest frame and the laser frequency in the lab frame. The current paper builds on a simplified approach suggested recently \cite{varcoe06} and presents a fully developed theory of the interaction within the Lorentz vilating electrodynamics of the Standard Model Extension \cite{kostelecky01}.
\end{abstract}
\maketitle

\section{Introduction}
Precision tests of special relativity are becomming more important as we continue to discover more about cosmology and particle physics. Dark matter, dark energy for example has a growing experimental support and little strong theoretical support. Theories that attempt to reconcile them with quantum electrodynamics (QED) and the standard model tend to introduce Lorentz violating features. Thus there is a growing suspician that a sufficently sensitive experiment will detect a violation. 
In an attempt to reconcile QED with potential violations of Lorentz invariance, a parameterisation of potential Lorentz violating terms has been attempted \cite{kostelecky01} in a Standard Model Extension (SME). 
The benefit of this parameterisation over previous versions is that it allows an experimental analysis to be built from Maxwell's equations. 
Many of the parameters in this formalism can be highly constrained by existing experiments, such as astronomical observations and accelerator tests. 
This leaves a few parameters that are both accessible to desktop optical experiments and cannot be easily constrained using other methods. 
The  experiments that have access to these parameters are the Michelson interferometer, the Kennedy-Thorndike test and the Ives-Stillwell test. 
Of these, the Ives-Stillwell test appears to be the only experimental test that is capable of accessing a parameter called $\kappa_{tr}$, and for this reason this is the least well known parameter of the SME. 

The Ives-Stillwell experiment is a precision measurement of the Doppler shift of light. 
The modern version of this experiment uses counter propagating lasers that are arranged to be co-linear with an accelerated ion beam. 
Tuning the lasers into resonance with the beam allows the lab frame frequency to be compared with the atomic rest frame frequency. Modern laser Ives-Stilwell experiments using storage-ring facilities have provided the tightest bounds on $\kappa_{tr}$ so far with an upper bound of $3 \times 10^{-8}$\cite{gwinner05}. Current predictions for future storage-ring experiments suggest that an upper bound of $\kappa_{tr} \le 1 \times 10^{-9}$ might be achievable in these techniques \cite{gwinnerpc}.  
The current paper explores the theory of a new version of the Ives-Stillwell experiment  using an EIT resonance in a sample of moving atoms as a potential tool for further improving this measurement.


In the framework of the SME, the sensitivity to an Ives-Stilwell experiment is given by the following formula \cite{tobar05}, 
\begin{equation}
\label{eq:is_1}
\dfrac{\nu_{E}\nu_{W}}{\nu_{0}^2} = 1 + 2\kappa_{tr}\left( \beta^{2} + 2\vec{\beta}_{\oplus}\cdot\vec{\beta} \right),
\end{equation}
where $\nu_{E}$ and $\nu_{W}$ are the Doppler shifted frequencies of the two counter propagating laser fields, equivalent to $\nu_{a}$ and $\nu_{p}$ only now defined to propagate along the axes of East and West in the laboratory frame: the orientation of the experiment in the laboratory frame is significant due to a sensitivity to the  Earth's velocity (labelling the fields in this manner removes any ambiguity between the direction of propagation and the coupling and probing fields required by EIT which will be encountered later). The rest frame transition frequency is $\nu_{0}$, $\kappa_{tr}$ is the SME parameter under investigation, $\vv=\beta c$ is the observer's velocity in the laboratory frame and $\beta_{\oplus}$ is the Earth's velocity in a frame centered upon the sun. It can be seen from equation \ref{eq:is_1} that there are two access channels to $\kappa_{tr}$ for experiments of this type, one relying solely on the observer's laboratory velocity $\beta$, the other on the vector product of the observer's velocity in the laboratory frame with the laboratories velocity in the Sun's frame $\beta_{\oplus}$. The sensitivity of the Ives-Stilwell experiments can be expressed as,
\begin{equation}
\dfrac{\nu_{E}\nu_{W}}{\nu_{0}^2} \simeq 1 + 2\kappa_{tr} \beta^{2}.
\label{eq:rein_simp}
\end{equation}
In these experiments the frequencies $\nu_{E}$, $\nu_{W}$ and $\nu_{0}$ are measured independently, this means that any uncertainty in frequency determination enters in quadrature. Although this can introduce significant systematic uncertainties, referencing to a frequency comb removes this as a limiting factor in these experiments. From equation (\ref{eq:rein_simp}), it is clear that the sensitivity of experiments scales quadratically with $\beta$, implying that increased velocities will lead to increased resolution. However, there is an element of diminishing returns associated with using extremely high observer velocities as relativity itself restricts the velocities attainable by virtue of $E^{2} = p^{2}c^{c} + m^2 c^4$. 

At present the sensitivity of Ives-Stilwell experiments is limited by the linewidth of the Lamb dip which is $10.8$MHz, about three times the natural linewidth of the transition. As a result of the relatively large linewidth current experiments are approaching the maximum sensitivity attainable by this detection scheme. 

One way to overcome the limitations imposed by relativity on the observers velocity, and saturation spectroscopy on the observed linewidth, is to move away from fast beams in favour of slower beams that enable one to create narrower linewidths.
In this paper we present a detailed experimental analysis of a new type of Ives-Stilwell apparatus which we will call the Modified Ives-Stilwell (MI) experiment. 
This takes a different approach to other experiments which are sensitive to a variation in the speed of light. 
However the concept is similar to the traditional Ives-Stillwell experiment in that a moving observer is used to measure the Doppler shifted frequency of two counter propagating laser fields. 
In this experiment a coherent interaction between the atom and the two laser fields creates an extremely narrow feature in the absorption spectrum. 
This can be orders of magnitude less than the excited state linewidth, and can be exploited as a precision frequency discriminator in a search for lorentz violating Doppler shift.

For the MI experiment, the optical fields counter propagate with respect to one another and the moving observer is a beam of $^{85}$Rb atoms. The two laser fields are chosen to have precisely the same laboratory frequency $\nu_{0}$.
\begin{figure}[ht!]
\centering
\psfrag{eps}{$\mathscr{L}$}
\psfrag{np}{$\nu_{0}$}
\psfrag{na}{$\nu_{0}$}
\psfrag{v}{$\vv_{at}$}
\subfigure[Laboratory frame:]{\includegraphics[width=6cm]{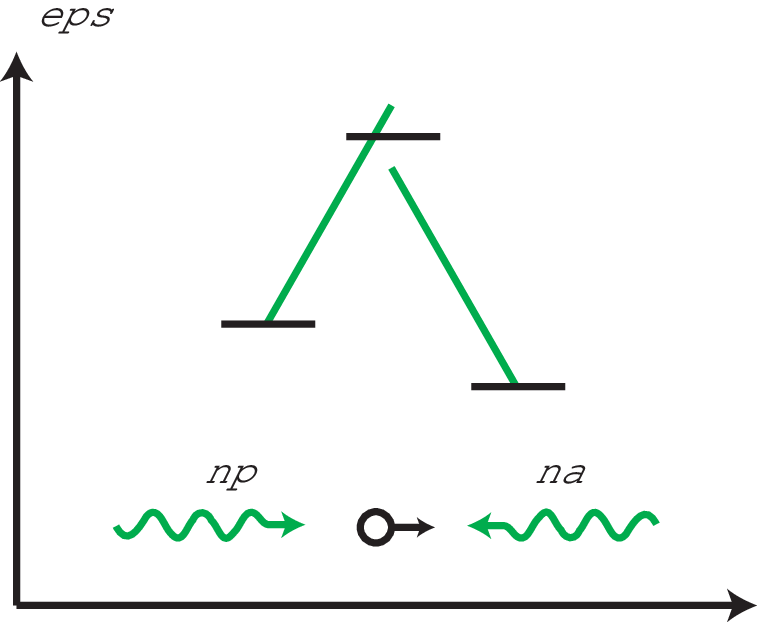}}
\hspace{1cm}
\psfrag{eps}{$\mathscr{E}$}
\psfrag{np}{$\nu_{p}$}
\psfrag{na}{$\nu_{a}$}
\subfigure[Atomic frame:]{\includegraphics[width=6cm]{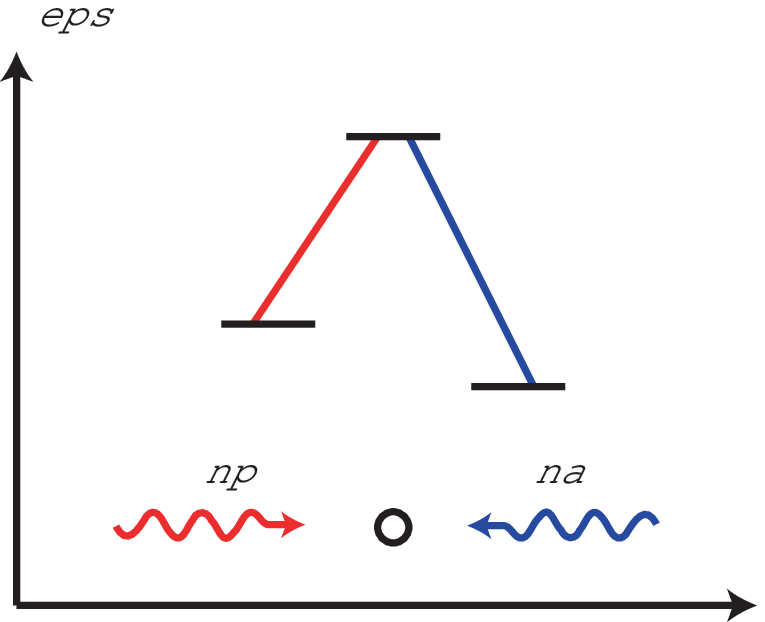}}
\caption[Experimental premise]{Experimental premise. In the rest frame of the laboratory the lasers used are far from resonance from either transition, however for atoms with an appropriate velocity both lasers are Doppler shifted into resonance forming a $\Lambda$-transition. A coherent interaction between the two optical fields and moving observer results in a coherent dark resonance in the absorption spectrum of one of the optical field which is extremely narrow. The line centre of the resultant coherent resonance is sensitive to small frame dependent changes in the speed of light. 
\label{fig:premise}}
\end{figure}

The co-linear alignment of the lasers is maintained by a folded interferometer and the laboratory frequency $\nu_{0}$ is chosen to be halfway between either leg of a $\Lambda$-system (see Figure \ref{fig:premise}). The $\lambda$ scheme is defined by the $F = 2\leftrightarrow F^{\prime}=2$ and $F = 3\leftrightarrow F^{\prime}=2$ hyperfine levels in the $5^{2}S_{1/2} \leftrightarrow 5^{2}P_{1/2}$, $D_{1}$ line, in $^{85}$Rb. The lab frame laser frequency is $\nu_{0} = 377.107,408,0THz$. 
The critical velocity is defined by the ground state hyperfine splitting for the $D_{1}$ line and is $\vv = 1207ms^{-1}$ for $^{85}$Rb and atoms with the critical velocity are trapped in a dark state by the combined interaction of the two laser fields.

The laser can be stabilized using an optical frequency comb and the linewidth of the coherent dark resonance we have achieved is less than $300$Hz in the current apparatus. 
Thus we are able to compare the lab frame fequency with the atomic frame frequency with high accuracy.

\section{The photon sector of the minimal SME}
In this section we breifly review the SME so that the terminology of the derivation that follows is placed in context. 
\kostelecky  and collaborators construct the Lorentz violating extension to QED by considering only CPT and Lorentz violating terms arising from the purely photon sector of the full SME. From this one arrives at the following Lagrangian \cite{colladay97,colladay98,kostelecky02}:
\begin{equation}
\label{eq:KL}
\mathscr{L}=-\frac{1}{4}F_{\mu\nu}F^{\mu\nu}+\frac{1}{2}\left(k_{AF}\right)^{\kappa}\epsilon_{\kappa\lambda\mu\nu}A^{\lambda}F^{\mu\nu}-\frac{1}{4}\left(k_{F}\right)_{\kappa\lambda\mu\nu}F^{\kappa\lambda}F^{\mu\nu}.
\end{equation}
In equation (\ref{eq:KL}) the electromagnetic field tensor is defined conventionally as $F_{\mu\nu}=(\partial_{\mu}A_{\nu} - \partial_{\nu}A_{\mu})$, while the coefficients $(k_{AF})^{\kappa}$ and  $(k_{F})_{\kappa\lambda\mu\nu}$ describe violations of the Lorentz and CPT symmetries respectively. In the limit that these coefficients are zero, equation (\ref{eq:KL}) reduces to the familiar QED Lagrangian in terms of $F_{\mu\nu}$ alone. The coefficient $(k_{F})_{\kappa\lambda\mu\nu}$ describes Lorentz violation. It shares all the symmetries of the Riemann tensor \cite{agacy99} and contains $19$ independent parameters \cite{kostelecky02}. The coefficient $(k_{AF})^{\kappa}$ is odd under CPT conjugation and has dimensions of mass.

There is an intimate relationship between the Lorentz and CPT symmetries; the construction of the CPT symmetry is Lorentz invariant. Hence Lorentz violation permits, but does not require CPT violation \cite{mattingly05}. Strong constraints have been placed on $(k_{AF})^{\kappa}$ via spectropolarimitary of distant galaxies \cite{carroll90, kostelecky01} and it has also been shown that finite values of $(k_{AF})^{\kappa}$ make the Lagrangian of equation (\ref{eq:KL}) unstable. It is therefore reasonable to ignore the CPT violating contributions to equation (\ref{eq:KL}). Consequently, setting $(k_{AF})^{\kappa} = 0$, greatly simplifies the Lagrangian of equation (\ref{eq:KL}), leavin a purely Lorentz violating contribution:
\begin{equation}
\label{eq:lag_1}
\mathscr{L}=-\frac{1}{4} F_{\mu\nu}F^{\mu\nu}-\frac{1}{4}\left(k_{F}\right)_{\kappa\lambda\mu\nu}F^{\kappa\lambda}F^{\mu\nu}.
\end{equation}
A physical interpretation of the tensor $(k_{F})^{\kappa\lambda\mu\nu}$ in equation (\ref{eq:lag_1}) containing non-zero terms is not obvious. However, there is a particularly useful decomposition of $(k_{F})^{\kappa\lambda\mu\nu}$ which can be used to present an analogy between \kostelecky's Lorentz violating extension to QED and the conventional condition of light propagating in homogeneous anisotropic media \cite{colladay97,colladay98,kostelecky02}:
\begin{align}
\label{eq:kap_decom1}
(\kappa_{DE})^{jk}&= -2(k_{F})^{0j0k}, \nonumber \\
(\kappa_{HB})^{jk}&= \frac{1}{2}\epsilon\indices{^j_{pq}}\epsilon\indices{^k_{rs}}(k_{F})^{pqrs}, \nonumber \\
(\kappa_{DB})^{jk}&= (k_{F})^{0jpq}\epsilon\indices{^k_{pq}}, \nonumber \\
(\kappa_{HE})^{jk}&= -(\kappa_{DB})^{kj}. 
\end{align} 
From these $3 \times 3$ matrices $(\kappa_{DE})^{jk}$, $(\kappa_{HB})^{jk}$, $(\kappa_{DB})^{jk}$ and $(\kappa_{HE})^{jk}$ a relationship between the electric displacement and auxiliary magnetic fields ${\bf D}$ and ${\bf H}$, and the electric and magnetic fields ${\bf E}$ and ${\bf B}$ can be derived via elements of the Lorentz violating parameter $(k_{F})^{\kappa\lambda\mu\nu}$ \cite{kostelecky02},
\begin{align}
\centering
\left(\begin{array}{c}\bf{D}\\\bf{H}\end{array}\right)&=\left(\begin{array}{cc}1+\kappa_{DE}&\kappa_{DB}\\\kappa_{HE}&1+\kappa_{HB}\end{array}\right)\left(\begin{array}{c}\bf{E}\\\bf{B}\end{array}\right).
\label{eq:kosteleckymatrix}
\end{align}
The Lagrangian of (\ref{eq:lag_1}) can then be expressed in terms of only the Lorentz violating parameters of equation (\ref{eq:kap_decom1}) and the electric and magnetic field components as \cite{kostelecky02}, 
\begin{equation}
\mathscr{L} = \dfrac{1}{2}\left( {\bf E}^2 - {\bf B}^2 \right) + \dfrac{1}{2}{\bf E}\cdot(\kappa_{DE})\cdot{\bf E} - \dfrac{1}{2}{\bf B}\cdot(\kappa_{HB})\cdot{\bf B} + {\bf E}\cdot(\kappa_{DB})\cdot{\bf B}.
\end{equation}
A further decomposition of $(k_{F})_{\kappa\lambda\mu\nu}$ defines a set of parameters which are directly accessible via experimentation \cite{kostelecky02}, 
\begin{align}
\label{eq:kap_oe}
 (\kappa_{e+})^{jk}&=\frac{1}{2}(\kappa_{DE} + \kappa_{HB})^{jk}, \nonumber \\
 (\kappa_{e-})^{jk}&=\frac{1}{2}(\kappa_{DE} - \kappa_{HB})^{jk} - \frac{1}{3}\delta^{jk}(\kappa_{DE})^{ll}, \nonumber \\
 (\kappa_{o+})^{jk}&=\frac{1}{2}(\kappa_{DB} + \kappa_{HE})^{jk}, \nonumber \\
 (\kappa_{o-})^{jk}&=\frac{1}{2}(\kappa_{DB} - \kappa_{HE})^{jk}, \nonumber \\
 \kappa_{tr}&=\frac{1}{3}(\kappa_{DE})^{ll}.
\end{align}
Each of the $\kappa$-parameters defined in equation (\ref{eq:kap_oe}) represents a $3\times3$ matrix except $\kappa_{tr}$ which is a scalar quantity. Parity odd terms are contained within the antisymmetric $\kappa_{o+}$ and symmetric $\kappa_{o-}$ and the parity even terms in the symmetric $\kappa_{e+}$, $\kappa_{e-}$ and $\kappa_{tr}$. Using the definitions given in equations (\ref{eq:kosteleckymatrix}) and equations (\ref{eq:kap_oe}) it can be shown that the  Lagrangian in equation (\ref{eq:lag_1}) can be re-expressed in such a way so as to make a direct connection between Lorentz violating parameters and experiments \cite{kostelecky02},
\begin{align}
\label{eq:lag_2}
\mathscr{L} &= \frac{1}{2}\left[ (1+\kappa_{tr}){\bf E}^{2}- (1-\kappa_{tr}){\bf B}^{2}\right]+\frac{1}{2}{\bf E}\cdot\left(\kappa_{e+}-\kappa_{e-}\right)\cdot{\bf E} \nonumber \\
~&- \frac{1}{2}{\bf B}\cdot\left(\kappa_{e+}-\kappa_{e-}\right)\cdot{\bf B}+{\bf E}\cdot\left(\kappa_{o+}-\kappa_{o-}\right)\cdot{\bf B}.
\end{align}
Inspection of equation (\ref{eq:lag_2}) immediately indicates a relationship between $\kappa_{tr}$ and the effective permittivity $\epsilon$ and permeability $\mu$,
\begin{equation}
(\epsilon - 1) = (\mu^{-1} + 1) = \kappa_{tr}.
\label{eq:sme_perm}
\end{equation}
Equation (\ref{eq:sme_perm}) can, albeit via an enormous oversimplification of the problem, be  related to the one-way speed of light,
\begin{equation}
\label{eq:c_ktr}
u = \frac{c}{1 + \kappa_{tr}}.
\end{equation}
Equation \ref{eq:c_ktr} illustrates how $\kappa_{tr}$ acts somewhat like a refractive index, though  in order to describe $u$ correctly requires an appropriate definition of reference frame in which $u$ is defined.

\subsection{The one-way speed of light}
\label{subsec:oneway}
The experiment described here, as well as conventional Ives-Stilwell experiments, are leading order sensitive to deviations in the one-way speed of light from the spacetime constant $c$. Although equation (\ref{eq:c_ktr}) illustrates a connection between $\kappa_{tr}$ and the speed of light, with $\kappa_{tr}$ acting somewhat like a refractive index. In order to corectly describe the experiment a full derivation in a Lorentz violating electrodynamics is required. This has been performed in a detailed analysis starting from the SME.

Using the definitions of the Lorentz violating electric displacement vector and auxiliary magnetic fields from equation (\ref{eq:kosteleckymatrix}) one can arrive at a Maxwell's equations, using the modified D and H fields from equation \ref{eq:kosteleckymatrix}, 
\begin{align}
\partial_{j}D^{j} &= 0, \\ 
\partial_{j}B^{j} &= 0, \\ 
\epsilon^{jkl}\partial_{k}H_{l} - \partial_{0}D^{j} &= 0, \label{eq:ampere} \\ 
\epsilon^{jkl}\partial_{k}B_{l} + \partial_{0}B^{j} &= 0.
\end{align}
From these, a modified dispersion relation for the electric field can be derived using the Lorentz violating analogue of Amp\`eres law, equation (\ref{eq:ampere}). Therefore, for an electromagnetic plane-wave with four-momentum $p^{\alpha}=(p^{0},p^{j})$, \cite{lan_ECM, kostelecky02}:
\begin{equation}
M^{jk}E^{k} = (-\delta^{jk} p^{2}  - p^{j} p^{k} - 2(k_{F})^{j \beta \gamma k}p_{\beta}p_{\gamma})E^{k}=0.
\label{eq:mod_amp}
\end{equation}
A condition of equation (\ref{eq:mod_amp}) is that the determinant of the coefficient should vanish, $|M^{jk}|=0$. Solving $|M^{jk}|=0$ one arrives at a relation between $p^{0}$ and $p^{j}$ \cite{kostelecky02},
\begin{equation}
\label{eq:dispersion}
p_{\pm}^{0}=(1 + \rho \pm \sigma)|p^{j}|,
\end{equation}
where $\rho=-\frac{1}{2}k\indices{_\alpha^\alpha}$, $\sigma^{2}=\frac{1}{2}(k_{\alpha\beta})^{2} - \rho^{2}$ and with,
\begin{equation}
k^{\alpha \beta}=(k_{F})^{\alpha\mu\beta\nu}\hat{p}_{\mu}\hat{p}_{\nu},
\label{eq:Thek}
\end{equation}
and $\hat{p}^{\mu}=p^{\mu}/|\vec{p}|$. 
From this dispersion relation the phase velocity of light, defined as $u=p_{0}p^{j}/\vec{p}^2$, can then be determined,
\begin{equation}
\label{eq:c_mod}
u=c(1+\rho\pm\sigma).
\end{equation}
This expression is important. It shows how, in this particular framework, deviations of the phase speed of light, $u$, from the spacetime constant $c$ are contained within the parameters $\rho$ and $\sigma$. The ability to describe an optical experiment in terms of the SME is  dependent on evaluating both $\rho$ and $\sigma$ in terms of parameters described in equation (\ref{eq:kap_oe}); this  requires a determination of $k^{\alpha \beta}$ in terms of these parameters also. Although $k^{\alpha \beta}$ must have been calculated by others previously, an explicit expression for it has never been published and so one is presented here. 

\begin{figure}
\centering
\psfrag{eps}{$\mathscr{E}$}
\psfrag{v}{$\vv_{at}$}
\psfrag{W}{West}
\psfrag{E}{East}
\psfrag{np}{$u_{W}$}
\psfrag{na}{$u_{E}$}
\psfrag{m1}{\small $\hat{p}^{\prime \mu} = (1;0,0,-1)$}
\psfrag{m2}{\small $\hat{p}^{\prime \mu} = (1;0,0,1)$}
\includegraphics[width = 10cm]{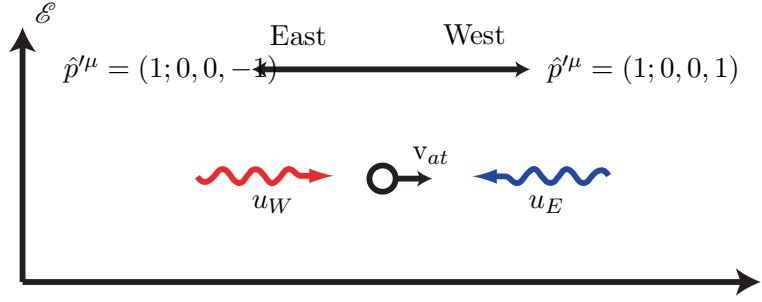}
\caption[The Experimental frame]{The Experimental frame, $\mathscr{E}$. Optical fields propagate  with phase speeds $u_{E}$ and $u_{W}$ along the unit four-vectors $\hat{p}^{\prime \mu} = (1;0,0,-1)$ and $\hat{p}^{\prime \mu} = (1;0,0,1)$ respectively.  The Easterly and Westerly unit vectors in a the laboratory frame line up with $\hat{p}^{\prime \mu} = (1;0,0,-1)$ and $\hat{p}^{\prime \mu} = (1;0,0,1)$.
\label{fig:exp_frame}}
\end{figure}

The tensor $k^{\alpha \beta}$ in equation (\ref{eq:Thek}) is a function of the wave unit-four-momentum, a consequence of which is that the terms $\rho$ and $\sigma$ from equation (\ref{eq:c_mod}) are both frame dependent. Therefore a meaningful evaluation of either $\rho$ or $\sigma$, and subsequently $u$, requires the specification of a reference frame. Although this reference frame can be freely chosen, calculations can be significantly simplified by choosing a special frame in which to evaluate $k^{\alpha \beta}$ and then making the appropriate transformation into an agreed frame where experiments are analysed and their results compared. For the case of an Ives-Stilwell experiment a suitable frame to choose is one at rest in the laboratory frame with it's third spatially axis aligned co-linearly with the two optical fields. In this special frame, which shall be refered to as the experimental frame and denoted by $\mathscr{E}$, the wave unit-four-vectors of the counter propagating fields therefore take the form $\hat{p}^{\prime\mu}=(1;0,0,\pm 1)$, see Figure \ref{fig:exp_frame}. 
Note that the third spatial component of $\hat{p}^{\prime\mu}$ to be $+1$ for fields propagating from East to West, and $-1$ for fields propagating from West to East throughout. 

The initial aim of this derivation was to be able to describe the phase speeds $u_{E}$ and $u_{W}$, for the two counter-propagating lasers used in this experiment in terms of the parameters $\kappa_{e+}$, $\kappa_{e-}$, $\kappa_{o+}$, $\kappa_{o-}$ and $\kappa_{tr}$, defined in equation (\ref{eq:kap_oe}), as this would then allow easy comparison of the results from this experiment with complimentary experiments. 

However, it is considerably easier and generally more useful to apply frame transformations to the `electrodynamic' parameters $\kappa_{DE}$, $\kappa_{HB}$, $\kappa_{DB}$ and $\kappa_{HE}$ defined in equation (\ref{eq:kap_decom1}). These parameters are explicitly defined in terms of $(k_{F})^{\alpha \mu \beta \nu}$ and are therefore more easily related to both $\rho$ and $\sigma$, a conversion into $\kappa_{e+}$, $\kappa_{e-}$, $\kappa_{o+}$, $\kappa_{o-}$ and $\kappa_{tr}$ via equation (\ref{eq:kap_oe}) then becomes possible. 


We begin the determination of $\rho$ and $\sigma$ by rearranging equations (\ref{eq:kap_decom1}) to give $(k_{F})^{\alpha\beta\mu\nu}$ terms of the matrices $\kappa_{DE}$, $\kappa_{HB}$, $\kappa_{DB}$ and $\kappa_{HE}$:
\begin{align}
(k_{F})^{0j0k} &= -\frac{1}{2}(\kappa_{DE})^{jk}, \nonumber \\
(k_{F})^{pqrs} &= \frac{1}{2}\epsilon\indices{_k^{rs}}\epsilon\indices{_j^{pq}}(\kappa_{HB})^{jk}, \nonumber \\
(k_{F})^{0jpq} &= \frac{1}{2}(\kappa_{DB})^{jk}\epsilon\indices{_k^{pq}}.
\end{align}
It should be noted that in order to completely describe $k_{F}$ using these equations it is extremely important to remember it has the same symmetry properties as the Riemann tensor\cite{agacy99} 

\begin{equation}
(k_{F})^{\alpha \beta \mu \nu} = -(k_{F})^{\beta \alpha \mu \nu} = -(k_{F})^{\alpha \beta \nu \mu} = (k_{F})^{\mu \nu \alpha \beta}
\end{equation}

and 

\begin{equation}
(k_{F})^{\alpha \beta \mu \nu} + (k_{F})^{\alpha \mu \nu \beta } + (k_{F})^{\alpha \nu \beta \mu} = 0
\end{equation}

Folowing this it is possible to fill all non-zero terms of in $(k_{F})^{\alpha \beta \mu \nu}$ with elements of $\kappa_{DE}$, $\kappa_{HB}$, $\kappa_{DB}$ and $\kappa_{HE}$. The tensor $(k_{F})^{\alpha \beta \mu \nu}$ could then be contracted with an appropriate wave unit-four-vector according to equation (\ref{eq:Thek}) to leave $k^{\alpha \beta}$ for both Easterly and Westerly propagating fields also in terms of $\kappa_{DE}$, $\kappa_{HB}$, $\kappa_{DB}$ and $\kappa_{HE}$ \cite{joethesis}. 
From which the Lorentz violating contributions to $u_{E}$ and $u_{W}$ can be determined.

With an expression for $k^{\alpha \beta}$ available in terms of $\kappa_{DE}$, $\kappa_{HB}$, $\kappa_{DB}$ and $\kappa_{HE}$ alone, and after evaluating $\rho$ in terms of $k^{\alpha \beta}$,
\begin{align}
\rho &= -\frac{1}{2}k\indices{_\alpha^\alpha} = -\frac{1}{2}\eta_{\alpha\beta}k^{\alpha\beta} \nonumber \\
 &= -\frac{1}{2}\left[ k^{00} -k^{11}-k^{22}-k^{33} \right],
\label{eq:rho1}
\end{align}
it was possible to rewrite equation (\ref{eq:rho1}) as for a beam propagating along the Easterly unit vector, $\hat{p}_{\mu}=(1;0,0,1)$, as:
\begin{equation}
\label{eq:rho2}
\rho = -\dfrac{1}{4}\left[ (\kappa_{DE} - \kappa_{HB})^{11} + (\kappa_{DE} - \kappa_{HB})^{22} + 2(\kappa_{DB})^{21} -2(\kappa_{DB})^{12} \right].
\end{equation}
To evaluate $\rho$ for $\hat{p}_{\mu}=(1;0,0,-1)$ simply change the sign of the $k_{DB}$ terms in equation (\ref{eq:rho2}).
Similarly, by evaluating $\sigma$ in terms of $k^{\alpha \beta}$,
\begin{align}
\sigma^2 & =  \frac{1}{2}(k_{\alpha\beta})^{2} - \rho^{2} \nonumber \\ ~ & =  [ (k^{00})^{2} + (k^{11})^{2} + (k^{22})^{2} + (k^{33})^{2} \nonumber \\ ~ & ~  + (k^{12})^{2} + (k^{13})^{2} + (k^{21})^{2} + (k^{23})^{2} + (k^{31})^{2} + (k^{32})^{2} \nonumber \\ ~ & ~  - (k^{01})^{2} - (k^{02})^{2} - (k^{03})^{2} - (k^{10})^{2} - (k^{20})^{2} - (k^{30})^{2} ]  \nonumber \\ ~ & ~ -\frac{1}{4}[ k^{00} -k^{11}-k^{22}-k^{33} ]^{2},
\label{eq:sig1}
\end{align}
and evaluating in terms of $\kappa_{DE}$, $\kappa_{HB}$, $\kappa_{DB}$ and $\kappa_{HE}$ for $\hat{p}_{\mu}=(1;0,0,1)$,
\begin{align}
\label{eq:sigma2}
\sigma^{2} &= \frac{1}{16} (4 [(k_{DB})^{11}]^2+4[ (k_{DB})^{12}]^2+8 (k_{DB})^{12} (k_{DB})^{21}+4[ (k_{DB})^{21}]^2-8 (k_{DB})^{11} (k_{DB})^{22} \nonumber \\
&+ 4 [(k_{DB})^{22}]^2-4 (k_{DB})^{12} (k_{DE})^{11}-4 (k_{DB})^{21} (k_{DE})^{11}+[(k_{DE})^{11}]^2+8 (k_{DB})^{11} (k_{DE})^{21}\nonumber \\
&- 8 (k_{DB})^{22} (k_{DE})^{21}+4 [(k_{DE})^{21}]^2 + 4 (k_{DB})^{12} (k_{DE})^{22}+4 (k_{DB})^{21} (k_{DE})^{22}\nonumber \\
&-2 (k_{DE})^{11} (k_{DE})^{22} +[(k_{DE})^{22}]^2-4 (k_{DB})^{12} (k_{HB})^{11}
-4 (k_{DB})^{21} (k_{HB})^{11}\nonumber \\
&+2 (k_{DE})^{11} (k_{HB})^{11}-2 (k_{DE})^{22} (k_{HB})^{11}  +[(k_{HB})^{11}]^2+8 (k_{DB})^{11} (k_{HB})^{21}\nonumber\\
&-8 (k_{DB})^{22} (k_{HB})^{21}+8 (k_{DE})^{21} (k_{HB})^{21}+4 [(k_{HB})^{21}]^2 +4 (k_{DB})^{12} (k_{HB})^{22}\nonumber \\
&+4 (k_{DB})^{21} (k_{HB})^{22}-2 (k_{DE})^{11} (k_{HB})^{22}+2 (k_{DE})^{22} (k_{HB})^{22} -2 (k_{HB})^{11} (k_{HB})^{22}+[(k_{HB})^{22}]^2 ).
\end{align}
Evaluating $\sigma$ is considerably more effort than $\rho$, but thankfully, it can be neglected for the purposes of this experiment. 

The tensors $k^{\alpha \beta}$ can be compared to the dielectric tensor of an anisotropic crystal from classical electrodynamics. Equations (\ref{eq:rho1}) and (\ref{eq:sig1}) show, $\rho$ and $\sigma$ are related to the trace and off-diagonal terms of $k^{\alpha \beta}$, which can be related to the phase speed and isotropy, and birefringent contributions from Lorentz violation respectively.

\section{EIT in a beam of $^{85}$Rb atoms}
The MI experiment seeks to measure any small deviations in the speed of light which manifest themselves as an asymmetric Doppler shift between two counter-propagating laser fields.  
In order to make any statement about the validity of the Doppler shift predicted by special relativity, knowledge of the observers precise velocity is required.
\subsection{A highly idealized system}
\label{sec:idealized}
Initially consider a beam of idealized three level atoms travelling with a velocity $\vv_{c}$ and interacting with two counter-propagating laser fields of equal rest frame frequency $\nu_{0}$ and orthogonal circular polarisations $\ket{\sigma_{\pm}}$. 
\begin{figure}[ht!]
\centering
{\small{
\subfigure[Two counter-propagating lasers interact with a three level atom travelling co-linearly with velocity $\vv_{c}$]{
\psfrag{V}{$\vv_{c}$}
\psfrag{m1}{$\ket{\Omega_{p},\sigma_{-}}$}
\psfrag{m4}{$\ket{\Omega_{c},\sigma_{+}}$}
\psfrag{m2}{~}
\psfrag{m3}{~}
\includegraphics[width=6cm]{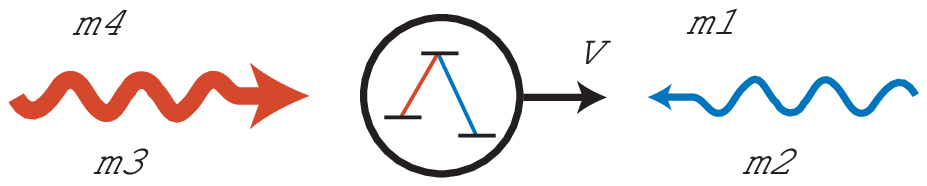}}
\hspace{0.5cm}
\subfigure[The laser fields form a $\Lambda$-system with the three level atom]{
\psfrag{1}{$\ket{m_{F}(1),g_{F}(1)}$}
\psfrag{2}{$\ket{m_{F}(2),g_{F}(2)}$}
\psfrag{3}{$\ket{m_{F}(3),g_{F}(3)}$}
\psfrag{f1}{$\ket{\Omega_{p},\sigma_{-}}$}
\psfrag{f2}{$\ket{\Omega_{c},\sigma_{+}}$}
\includegraphics[width=6cm]{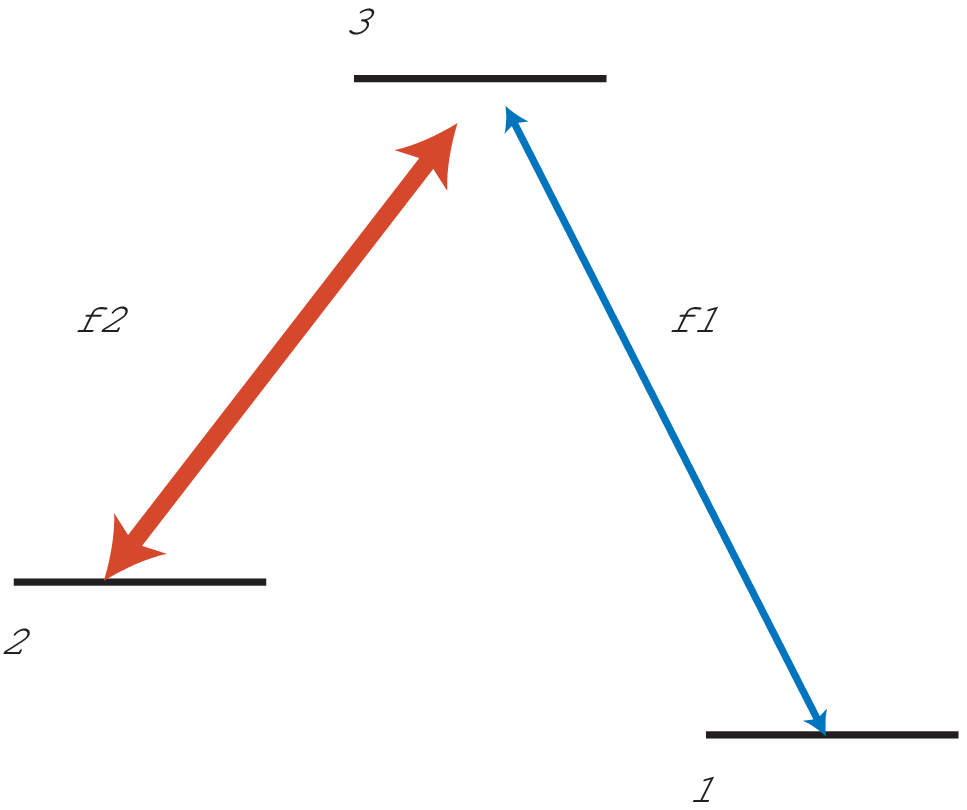}}
}}
\caption[EIT in idealized three level beam]{EIT in idealized three level beam. The lasers are assumed to have equal rest frame frequency $\nu_{0}$ chosen such that for an atomic velocity $\vv_{c}$ the weak probe field and strong coupling field with orthogonal circular polarisations form a $\Lambda$-system with the three energy levels of the atom. 
\label{fig:idealEIT1}}
\end{figure}
The frequency $\nu_{0}$ is chosen such that the approaching field will be blue shifted into resonance with the $\nu_{31}$ transition while the receeding field will be red shifted into resonance with the $\nu_{32}$ transition. For a symmetric Doppler shift of the two fields this  occurs for a laser frequency of,
\begin{equation}
\nu_{0} = \dfrac{\nu_{31} + \nu_{32}}{2},
\end{equation}
in an atom travelling with velocity $\vv_{c}$ defined as,
\begin{equation}
\label{eq:velocity}
\vv_{c} = \dfrac{\nu_{21}}{2\nu_{0}}c = \dfrac{\nu_{31} - \nu_{32}}{\nu_{31} + \nu_{32}}c.
\end{equation}

If the states $\ket{1}$, $\ket{2}$ and $\ket{3}$ of our moving observer are composed of Zeeman substates with magnetic quantum numbers $m_{F}(j)$ and gyromagnetic ratios $g_{F}(j)$ for state $\ket{j}$ and $|m_{F}(3) - m_{F}(2)|=|m_{F}(3) - m_{F}(1)| = 1$ and $|m_{F}(2) - m_{F}(1)| = 2$ then the three levels can form a $\Lambda$ system using orthogonal circular polarisations $\sigma_{+}$ and $\sigma_{-}$, see Figure \ref{fig:idealEIT1}. In this model the field co-propagating with the atomic beam is strong and therefore considered a coupling field, while the field counter-propagating with the atomic beam is considered a probing field. Zeeman sub levels are sensitive to magnetic fields. The change in frequency, $\Delta\nu_{B}$, of such a state induced by an applied magnetic field is described by \cite{footbook}, 
\begin{equation}
\Delta\nu_{B} = \frac{\mu_{B}}{h}m_{F}g_{F}B,
\end{equation}
where B is the magnetic field along the axis of the interaction. An externally applied magnetic field can therefore contribute to the detunings $\Delta_{1}$ and $Delta_{2}$.

For an exquisitely defined atomic beam, i.e. an atomic beam with a velocity distribution approximating a delta function, $\delta(\vv - \vv_{c})$, and considering only the first order contribution to the Doppler shift, the detunings $\Delta$ and $\delta$ are therefore described by the formulae:
\begin{align}
\label{eq:EIT_detunings}
\Delta(B,\vv_{c}) &=  \nu_{31}(0) + \frac{\mu_{B}}{h}\left[m_{F}(3)g_{F}(3) - m_{F}(1)g_{F}(1)\right]B - \nu_{0}(1 + \vv_{c}/c), \nonumber \\
\delta(B,\vv_{c}) &=  \nu_{21}(0) + \frac{\mu_{B}}{h}\left[m_{F}(2)g_{F}(2) - m_{F}(1)g_{F}(1)\right]B -2 \nu_{0}\vv_{c}/c,
\end{align}
where $\nu_{jk}(0)$ represents the frequency splitting between $\ket{j}\leftrightarrow\ket{k}$ for $B = 0$. These detunings provides a description of the absorption as a function of applied magnetic field.
\begin{figure}[ht!]
\centering
%
%
\begin{psfrags}%
\psfragscanon%
%
\psfrag{s03}[t][t]{\color[rgb]{0,0,0}\setlength{\tabcolsep}{0pt}\begin{tabular}{c}B/nT\end{tabular}}%
\psfrag{s04}[b][b]{\color[rgb]{0,0,0}\setlength{\tabcolsep}{0pt}\begin{tabular}{c}Normalised absorption / arb. un.\end{tabular}}%
%
\psfrag{x01}[t][t]{$-40$}%
\psfrag{x02}[t][t]{$-20$}%
\psfrag{x03}[t][t]{$0$}%
\psfrag{x04}[t][t]{$20$}%
\psfrag{x05}[t][t]{$40$}
%
\psfrag{v01}[r][r]{~}%
\psfrag{v02}[r][r]{$0.2$}%
\psfrag{v03}[r][r]{~}%
\psfrag{v04}[r][r]{$0.4$}%
\psfrag{v05}[r][r]{~}%
\psfrag{v06}[r][r]{$0.6$}%
\psfrag{v07}[r][r]{~}%
\psfrag{v08}[r][r]{$0.8$}%
\psfrag{v09}[r][r]{~}%
%
\includegraphics[width=10cm]{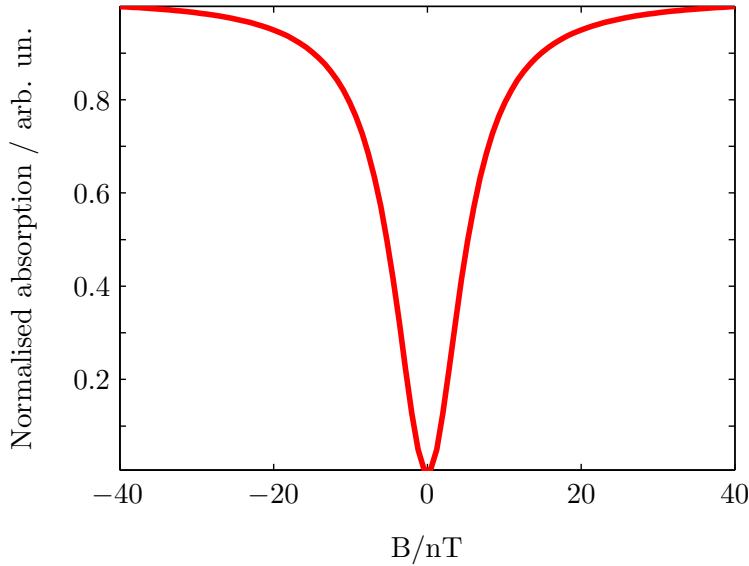}%
\end{psfrags}%
%

\caption[EIT in a hypothetical atomic beam]{The normalized EIT imaginary susceptibility for a hypothetical atomic beam. The beam is exquisitely defined with velocity $\vv_{c}$ and the coupling and probing fields are counter-propagating with respect to one another. See text for a description of the parameters used in this simulation. 
\label{fig:chiEITbeam}}
\end{figure}
Figure \ref{fig:chiEITbeam} shows the effects of Zeeman detunings in an exquisitely defined atomic beam as described previously. The parameters used are based on a three level system with the same hyperfine structure as $^{85}$Rb: $\nu_{31} = 377,108,911.7$MHz, $\nu_{32} = 377,105,876.0$MHz and $\nu_{21} = 3035.732$MHz \cite{barwood91}. The states are $\ket{1} = \ket{F = 2, g_{F} = -1/9,m_{F} = 2}$, $\ket{2} = \ket{F = 3, g_{F} = 1/9,m_{F} = 0}$ and $\ket{3} = \ket{F^{\prime}=2, g_{F} = -1/3,m_{F} = 1}$. The dephasing terms used were $\gamma_{31} = 1\times10^{11}$Hz and $\gamma_{21} = 0$Hz and the coupling field took a value of $0.0001\gamma_{31}$. The velocity $\vv_{c} = (\nu_{31} - \nu_{32})/(\nu_{31} + \nu_{32}) \sim 1207ms^{-1}$. 


\subsection{Doppler broadening}
An exquisitely defined velocity distribution is of course highly unrealistic as the velocity distribution will governed by Maxwell-Boltzmann statistics. Subsequently, there will be a large number of atoms of varying velocities for the coupling and probing fields to interact with. With the Doppler width of rubidium being some hundreds of MHz one might expect the narrow transmission window depicted in Figure \ref{fig:chiEITbeam} to be washed out. However, numerical simulations have shown this not to be the case.
\begin{figure}[ht!]
\centering
\psfrag{n}{$\nu$}
\psfrag{n32}{$\nu_{32}$}
\psfrag{n31}{$\nu_{31}$}
\psfrag{vc}{$v_{c}$}
\subfigure[Exactly on resonance: The atom has precisely the critical velocity $v_{c}$ such that the parallel and antiparallel laser fields are Doppler shifted into resonance with the transitions $\nu_{32}$ and $\nu_{31}$ respectively.]{\includegraphics[width = 10cm]{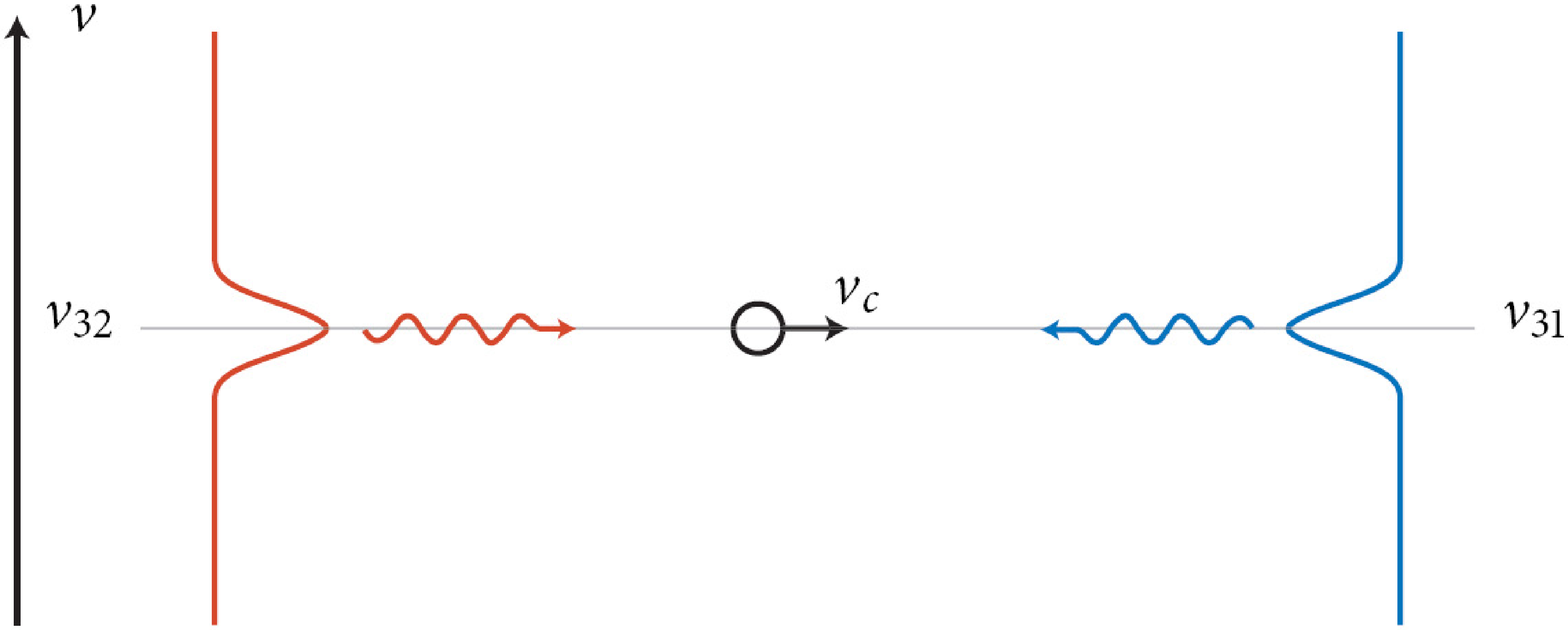}}
\psfrag{n}{$\nu$}
\psfrag{n32}{$\nu_{32}$}
\psfrag{n31}{$\nu_{31}$}
\psfrag{vc}{$v_{c} + \delta v$}
\subfigure[Small velocity detuning: The atomic velocity is slightly detuned from  $v_{c}$ such that the resonance frequencies $\nu_{32}$ falls on the red detuned side of the laser/excited state linewidth and the resonance frequencies $\nu_{31}$ falls on the blue detuned side of the laser/excited state linewidth.]{\includegraphics[width = 10cm]{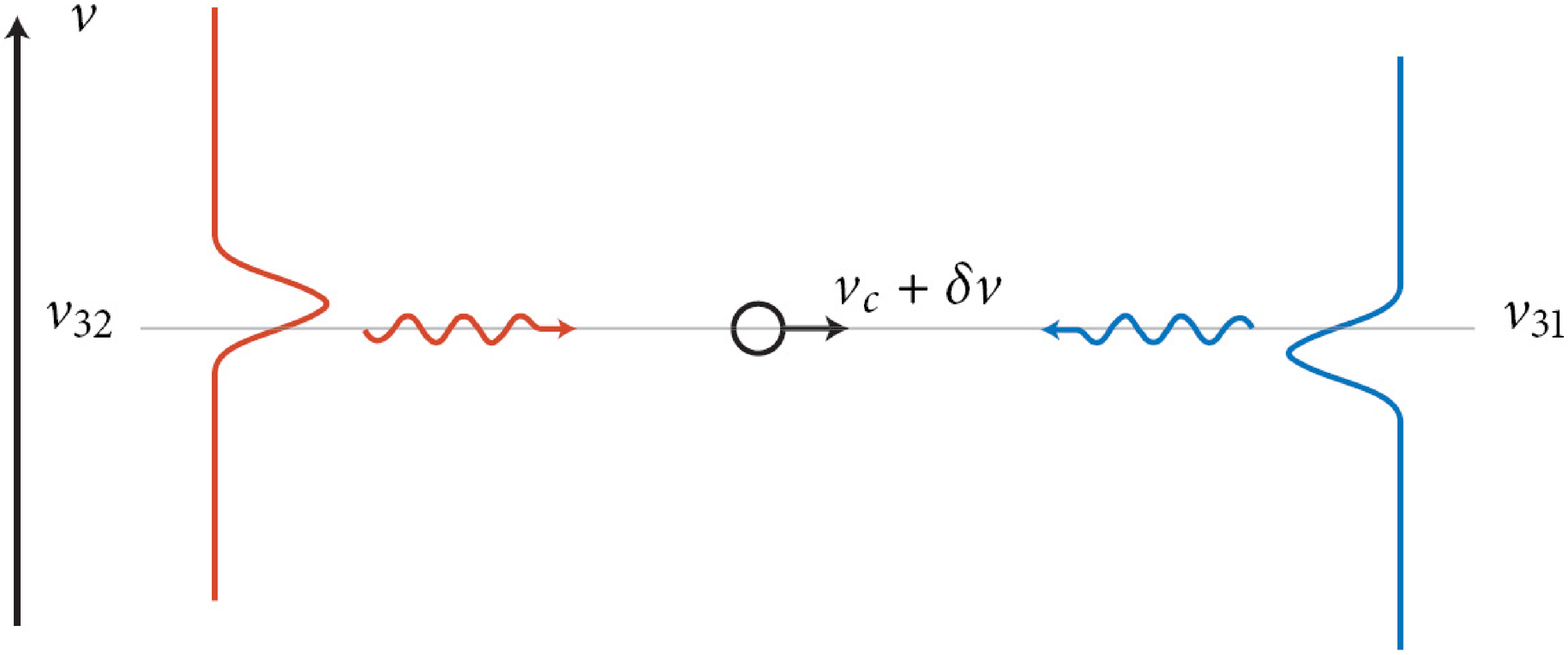}}
\psfrag{n}{$\nu$}
\psfrag{n32}{$\nu_{32}$}
\psfrag{n31}{$\nu_{31}$}
\psfrag{vc}{$v_{c} + \delta v$}
\subfigure[Large velocity detuning: For atomic velocities far from the critical velocity $v_{c}$ the Doppler shift is so great that neither frequency is anywhere near the resonance frequencies $\nu_{32}$ or $\nu_{31}$.]{\includegraphics[width = 10cm]{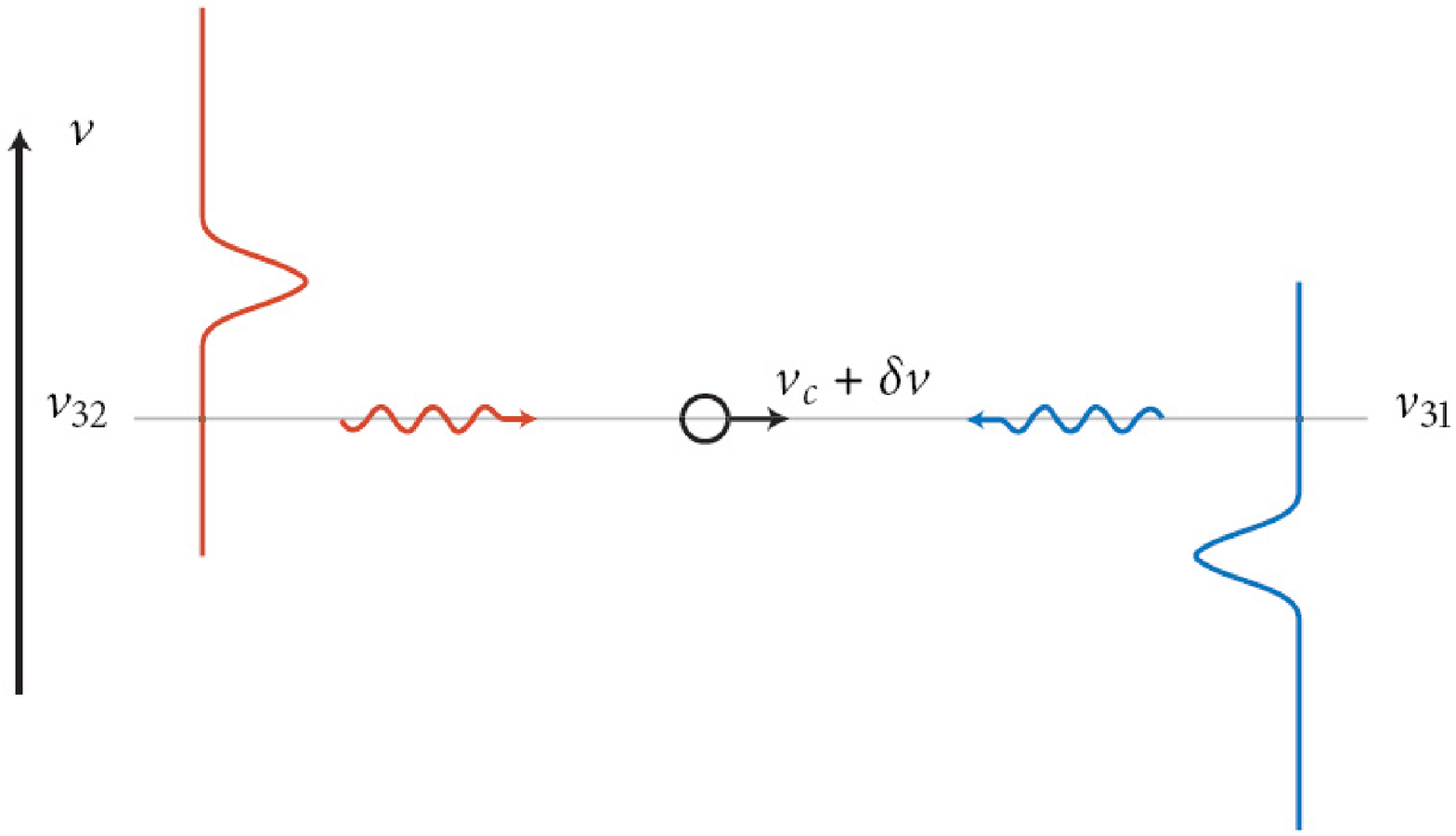}}
\caption[Why this system is velocity selective]{Why this system is velocity selective. Both lasers have the same rest frame frequency $\nu_{0}$, so an atom with velocity $\vv \ne \vv_{c}$ will see an equal and opposite Doppler shift away from resonance for the two optical fields. Therefore, atoms travelling along the experimental axis with velocity $\vv_{c}$ will satisfy the two photon Raman resonance required for EIT. However, small deviations in the atomic velocity from this  critical value $\vv_{c}$ prevent this Raman resonance occurring and subsequently EIT cannot be generated. The red and blue peaks represent the transition linewidth of the excited state.
\label{fig:v_selection}}
\end{figure}

The Maxwell-Boltzmann velocity vector distribution function is described by,
\begin{equation} 
P(\vv_{j}) = \sqrt{\dfrac{m}{2\pi k T}}\exp\left( - \dfrac{m \vv_{j}^{2}}{2kT} \right).
\end{equation}
This describes the probability, $P(\vv_{j})$, of an atom travelling along the unit vector $\hat{j}$ with a speed $\vv_{j}$, $T$ is the absolute temperature, $k$ is the Boltzmann constant and $m$ is the atomic mass. In  the cell there will be a number of atoms of different velocities and therefore detunings, $\Delta$ and $\delta$, contributing to the overall observed susceptibility. This can be simulated by integrating the product of the imaginary part of the susceptibility with the velocity distribution function,
\begin{align}
I &= \int \Im\left[\chi^{(1)}(B,\vv_{j})\right]P(\vv_{j})d\vv_{j}.
\label{eq:int_1}
\end{align}
The critical velocity $\vv_{c} = 1207ms^{-1}$ for the $D_{1}$ line of rubidium is far out in the tail of the Gaussian velocity distribution where the distribution is quite flat. Because both lasers are derived from the same source, and therefore have the same $\nu_{0}$, an atom with velocity $\vv \ne \vv_{c}$ will see an equal and opposite Doppler shift away from resonance for the two optical fields. 

Coherent phenomena such as EIT or coherent population trapping have previously been shown to exhibit velocity selective behaviour \cite{aspect88}. A simple argument to explain the velocity selective nature of EIT in our scheme is depicted in Figure \ref{fig:v_selection}. For an atomic beam travelling with velocity $\vv_{c}$ the two photon Raman resonance condition is satisfied and EIT can occur. However, for small deviations from this critical velocity the Raman resonance is not fulfilled by virtue of opposite Doppler shifts in the two interacting beams, subsequently EIT cannot be generated. 

\begin{figure}[ht!]
\centering
\includegraphics{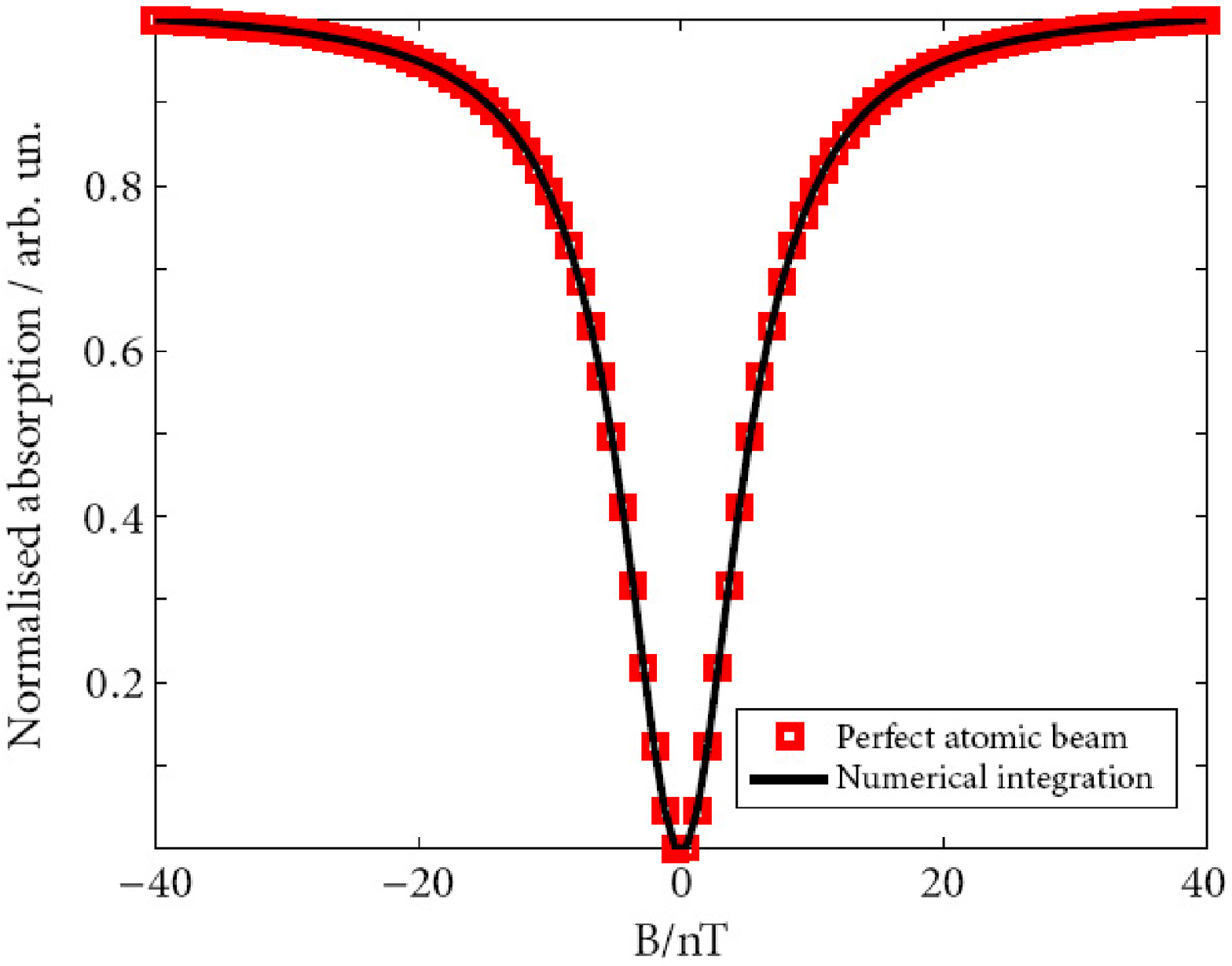}
\caption[Numerical integration of Doppler broadening]{Numerical integration of Doppler broadening. This figure compares the imaginary susceptibility for that of an exquisitely defined  atomic beam (red) and a thermal distribution after numerical integration of the atomic velocity from $\vv_{c}\pm4.4ms^{-1}$ (black). The integration limit was defined by the excited state linewidth of $5.746$MHz. Both spectra have been normalized to their own maximum value. See text for further details.
\label{fig:intV}}
\end{figure}

In addition to this qualitative argument, a numerical integration of equation (\ref{eq:int_1}) was performed for one hundred points ranging from $-40\rightarrow40$nT between the limits of $\pm4.4ms^{-1}$ , see Figure \ref{fig:intV}. These limits are defined by the excited state linewidth which is $5.746$MHz.
The lineshape which results from this numerical integration is indistinguishable from that of an exquisitely defined atomic beam of velocity $\vv_{c}$. It is as though the velocity distribution  has collapsed into a delta function making the transition velocity selective; only those atoms with a velocity close to $\vv_{c}$ contribute to the observed signal. For a bariety of conditions, the resultant lineshape remained unaffected for integration limits up to several hundred MHz. The transition is therefore extremely insensitive to Doppler broadening.

\section{The Ives-Stilwell experiment and the SME}
To analyse the MI experiment in the SME we begin by defining the Doppler shifted frequencies of the two counter-propagating fields in the reference frame of the atom. For an experiment as depicted in Figure \ref{fig:exp_frame} and including any modifications to the phase velocity which might occur from a violation of Lorentz invariance, i.e. $u_{E,W}$, the Doppler shift can be described as,
Where the phase velocity of light is given by equation \ref{eq:c_mod} and $\nu_E$ and $\nu_{W}$  refers to the frequency of light seen by an observer traveling along an Easterly and Westerly unit vector respectively, with an observer also propagating along the westerly unit vector.

\subsection{The Modified Ives-Stilwell experiment}
In equations \ref{eq:EIT_detunings} we showed that the one and two photon detunings, $\Delta$ and $\delta$, can be described by, 
\begin{align}
\label{eq:SME-detunings}
\Delta(B,\vv_{c}) &=  \nu_{31}(0) + \frac{\mu_{B}}{h}\left[m_{F}(3)g_{F}(3) - m_{F}(1)g_{F}(1)\right]B - \nu_{0}(1 + \vv_{c}/c), \nonumber \\
\delta(B,\vv_{c}) &=  \nu_{21}(0) + \frac{\mu_{B}}{h}\left[m_{F}(2)g_{F}(2) - m_{F}(1)g_{F}(1)\right]B -2 \nu_{0}\vv_{c}/c.
\end{align}
However, these detunings do not take into account any Lorentz violating contributions. In order to do so, the spacetime constant $c$ appearing in equation (\ref{eq:SME-detunings}), which arrises from the Doppler shift, must be replaced by the appropriate phase speed $u_{E,W}$ from equation (\ref{eq:c_mod}). For an atomic beam propagating Easterly with a probe field propagating Westerly, and expanding only to leading order in atomic velocity $\beta_{at}$ and $\rho$, the one and two photon detunings $\Delta$ and $\delta$ can be written as:
\begin{align}
\Delta(B,\beta_{at}) &=  \nu_{31}(0) + \frac{\mu_{B}}{h}\left[m_{F}(3)g_{F}(3) - m_{F}(1)g_{F}(1)\right]B - \nu_{0}(1 + \beta_{at}) + \nu_{0}\beta_{at}\rho_{E}, \nonumber \\
&= \nu_{31}(B) - \nu_{0}(1 + \beta_{at}) + \nu_{0}\beta_{at}\rho_{E}, \label{eq:detunings_1a} \\ \nonumber \\
\delta(B,\beta_{at}) &=  \nu_{21}(0) + \frac{\mu_{B}}{h}\left[m_{F}(2)g_{F}(2) - m_{F}(1)g_{F}(1)\right]B  -2\nu_{0}\beta_{at} + \nu_{0}\beta_{at}(\rho_{E} + \rho_{W}), \nonumber \\
&= \nu_{21}(B) -2\nu_{0}\beta_{at} + \nu_{0}\beta_{at}(\rho_{E} + \rho_{W}), \nonumber \\
&= \nu_{21}(B) -2\nu_{0}\beta_{at} - 2\kappa_{tr}\nu_{0}\beta_{at} \label{eq:detunings_1b} ,
\end{align}
where $\beta_{at} = \vv_{at}/c$. The final terms in equation (\ref{eq:detunings_1a}) and (\ref{eq:detunings_1b}) contain the Lorentz violating contributions to the detunings. As one would hope, in the limit that $\kappa_{tr} = 0$ these detunings reduce to those predicted by special relativity. 

Simulations of the expected EIT signal for finite values of $\kappa_{tr}$ revealed it to be far less sensitive to a Lorentz violating contribution to the single photon detunings, $\Delta$, than it is two the photon detunings, $\delta$. These simulations demonstrated that the effect of even large values of $\kappa_{tr}$ in this context had little effect on the observed spectra, with noticeable effects for the current experimental resolution only arising for values of $\kappa_{tr}$ orders of magnitude larger than the current upper bound, $\kappa_{tr} \le 8.3 \times 10^{-8}$ \cite{reinhardt07}.   

In contrast, the contribution of $\kappa_{tr}$ to the observed EIT spectra from the two-photon detuning $\delta$ was very different, having an enormous effect on the magnetic field at which the peak of the spectrum occured. This effect persisted even after integration over the thermal velocity distribution, confirming that the resonance remains velocity selective even in the presence of Lorentz violation. 

From equation (\ref{eq:detunings_1b}) it can be seen that the offset of the peak, $\Xi$, arrising solely from Lorentz violating effects, for such an experiment is:
\begin{align}
\label{eq:off_freq}
\Xi &= 2\kappa_{tr} \nu_{0}\beta_{at}, \nonumber \\
~ &\sim 3 \times 10^{9} \kappa_{tr} Hz.
\end{align}
\begin{figure}[ht!]
\centering
%
%
\begin{psfrags}%
\psfragscanon%
%
\psfrag{s01}[t][t]{\color[rgb]{0,0,0}\setlength{\tabcolsep}{0pt}\begin{tabular}{c}$\Xi$\end{tabular}}%
\psfrag{s02}[b][b]{\color[rgb]{0,0,0}\setlength{\tabcolsep}{0pt}\begin{tabular}{c}Absorption / arb. un.\end{tabular}}%
\psfrag{s03}[lB][lB]{\color[rgb]{0,0,0}\setlength{\tabcolsep}{0pt}\begin{tabular}{l}$\Xi$\end{tabular}}%
%
\psfrag{x01}[t][t]{$0$}%
\psfrag{x02}[t][t]{~}%
\psfrag{x03}[t][t]{~}%
\psfrag{x04}[t][t]{~}%
\psfrag{x05}[t][t]{~}%
\psfrag{x06}[t][t]{$0.5$}%
\psfrag{x07}[t][t]{~}%
\psfrag{x08}[t][t]{~}%
\psfrag{x09}[t][t]{~}%
\psfrag{x10}[t][t]{~}%
\psfrag{x11}[t][t]{$1$}%
\psfrag{x12}[t][t]{~}%
\psfrag{x13}[t][t]{~}%
\psfrag{x14}[t][t]{$0$}%
\psfrag{x15}[t][t]{~}%
\psfrag{x16}[t][t]{$1$}%
\psfrag{x17}[t][t]{~}%
\psfrag{x18}[t][t]{~}%
%
\psfrag{v01}[r][r]{$0$}%
\psfrag{v02}[r][r]{~}%
\psfrag{v03}[r][r]{~}%
\psfrag{v04}[r][r]{~}%
\psfrag{v05}[r][r]{~}%
\psfrag{v06}[r][r]{$0.5$}%
\psfrag{v07}[r][r]{~}%
\psfrag{v08}[r][r]{~}%
\psfrag{v09}[r][r]{~}%
\psfrag{v10}[r][r]{~}%
\psfrag{v11}[r][r]{$1$}%
\psfrag{v12}[r][r]{$0$}%
\psfrag{v13}[r][r]{~}%
\psfrag{v14}[r][r]{~}%
\psfrag{v15}[r][r]{~}%
\psfrag{v16}[r][r]{~}%
\psfrag{v17}[r][r]{~}%
\psfrag{v18}[r][r]{~}%
\psfrag{v19}[r][r]{~}%
\psfrag{v20}[r][r]{~}%
\psfrag{v21}[r][r]{~}%
\psfrag{v22}[r][r]{$1$}%
%
\includegraphics[width=10cm]{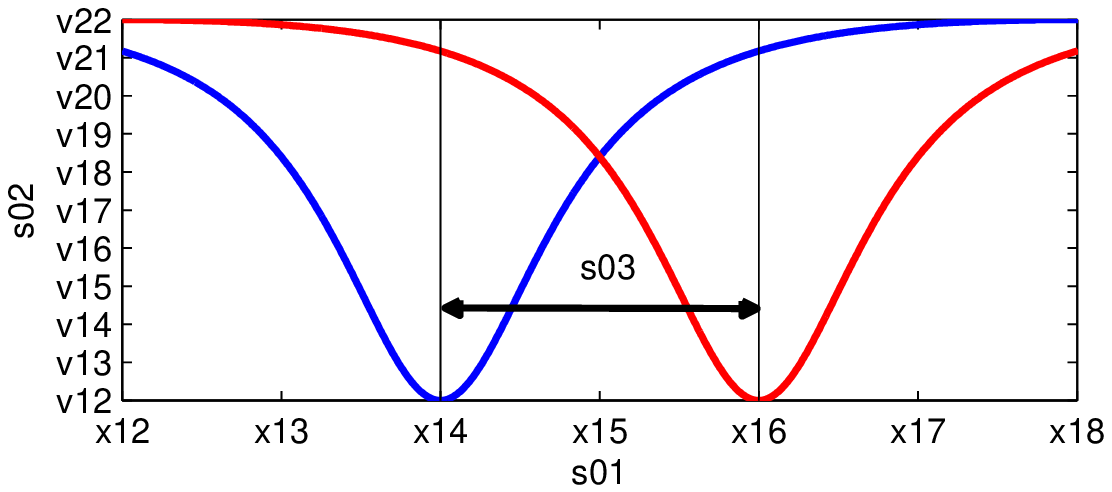}%
\end{psfrags}%
%

\caption[$\kappa_{tr}$ modification to EIT spectrum]{A finite value of $\kappa_{tr}$ shifts the EIT line centre from zero magnetic field to a value $\Xi = 2\kappa_{tr}\nu_{0}\beta_{at}$. The blue line represents the spectrum in the absence of Lorentz violation, the red line represents the spectrum for a finite value of $\kappa_{tr}$.
\label{fig:splitting_pred}}
\end{figure}

For the most magnetically sensitive transition in the $D1$ line of $^{85}$Rb an external magnetic field will cause a change in the two photon detuning of $\sim 1MHz/Gs = 10^{10}Hz/Tesla$, so the offset described by equation (\ref{eq:off_freq}) is approximately equivalent to:
\begin{align}
\label{eq:off_mag}
\Xi &= 2\kappa_{tr} \nu_{0}\beta_{at}, \nonumber \\
~ &\sim 0.3 \kappa_{tr} Tesla.
\end{align}
For a value of $\kappa_{tr}$ consistent with the current experimental upper bound this would correspond to a shift of $25$nT in the peak of an EIT signal. Magnetometry at this level of sensitivity is already achievable, with magnetometers relying on phenomona such as EIT being shown to exhibit sensitivities already on the order of $10pT/\sqrt{Hz}$ \cite{belfi07pp,bison04}  

Although an obvious experiment to measure $\kappa_{tr}$ might be to look for the deviations of an EIT resonance line centre from $0$T magnetic field in an atomic beam. There are a number of factors, other than Lorentz violation, that could cause a shift in the line centre of such a transition, with stray magnetic fields being of particular concern. Such a measurement would be fraught with systematic uncertainties. An alternative, which is capable of removing many of these systematic effects, is to make a differential measurement. 

\begin{figure}[ht!]
\centering
\psfrag{DD}{\small $\Delta_{1}$}
\psfrag{D1}{\small $\Delta_{2}$}
\psfrag{G31}{\small $\Gamma_{31}$}
\psfrag{G32}{\small $\Gamma_{32}$}
\psfrag{1}{\small $\ket{1}$}
\psfrag{2}{\small $\ket{2}$}
\psfrag{3}{\small $\ket{3}$}
\psfrag{n31}{\small $\nu_{31}$}
\psfrag{n32}{\small $\nu_{32}$}
\psfrag{n21}{\small $\nu_{21}$}
\psfrag{np}{\small $\nu_{p}$}
\psfrag{nc}{\small $\nu_{c}$}
\psfrag{E}{\small West}
\psfrag{W}{\small East}
\includegraphics[width = 14cm]{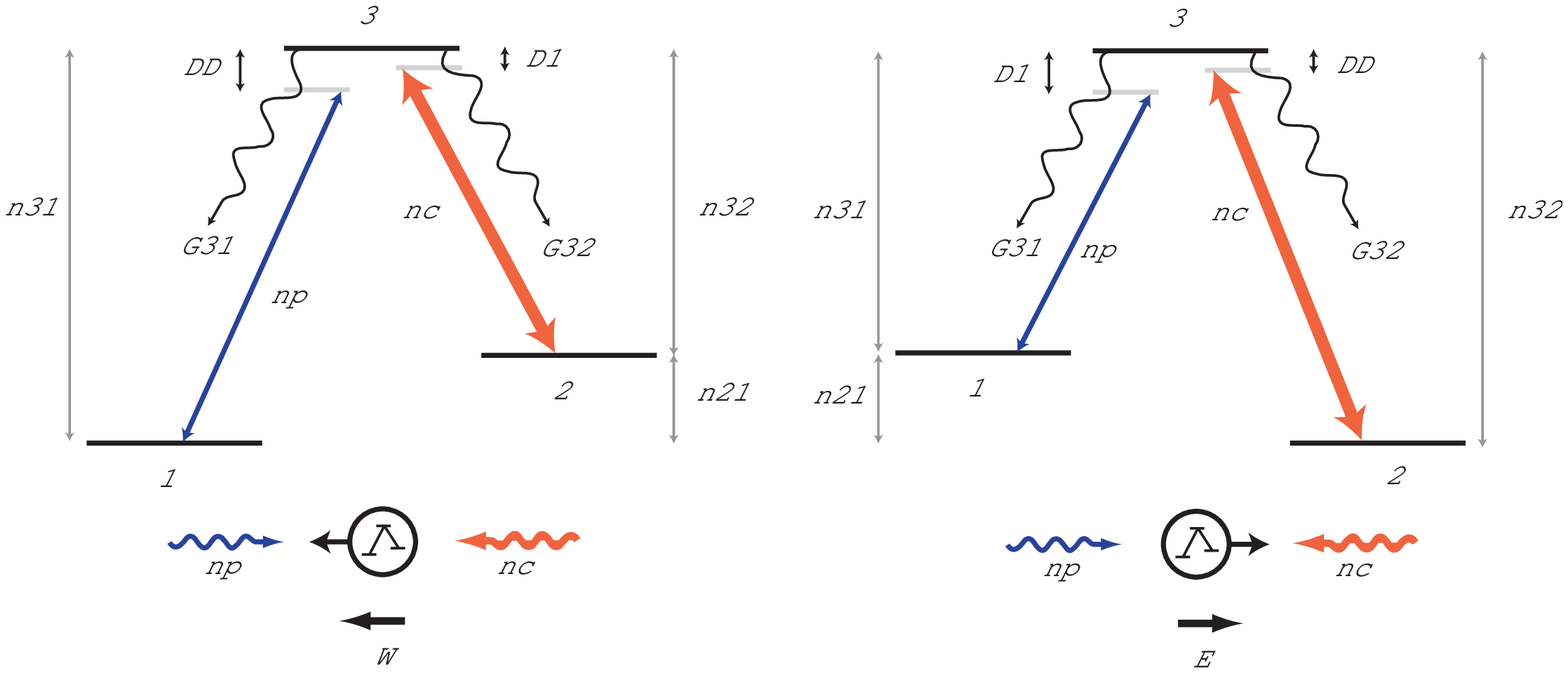}
\caption[Toy-model of two counter propagating $\Lambda$-systems]{Toy-model of two counter propagating $\Lambda$-systems. Under the influence of a finite value of $\kappa_{tr}$ the resonance condition of these two systems shifts in opposite directions. 
\label{fig:toymodel}}
\end{figure}

For the model considered here, there should be two atomic beams propagating in opposite directions along the experimental axis which could support an EIT resonance. 
\begin{figure}[ht!]
\begin{center}
\psfrag{X}{$\hat{x}$}
\psfrag{Y}{$\hat{y}$}
\psfrag{Z}{$\hat{z}$}
\psfrag{E}{East}
\psfrag{W}{West}
\psfrag{nc}{$\nu_{c}$}
\psfrag{np}{$\nu_{p}$}
\psfrag{b}{$\beta_{at}$}
\includegraphics[width=12cm]{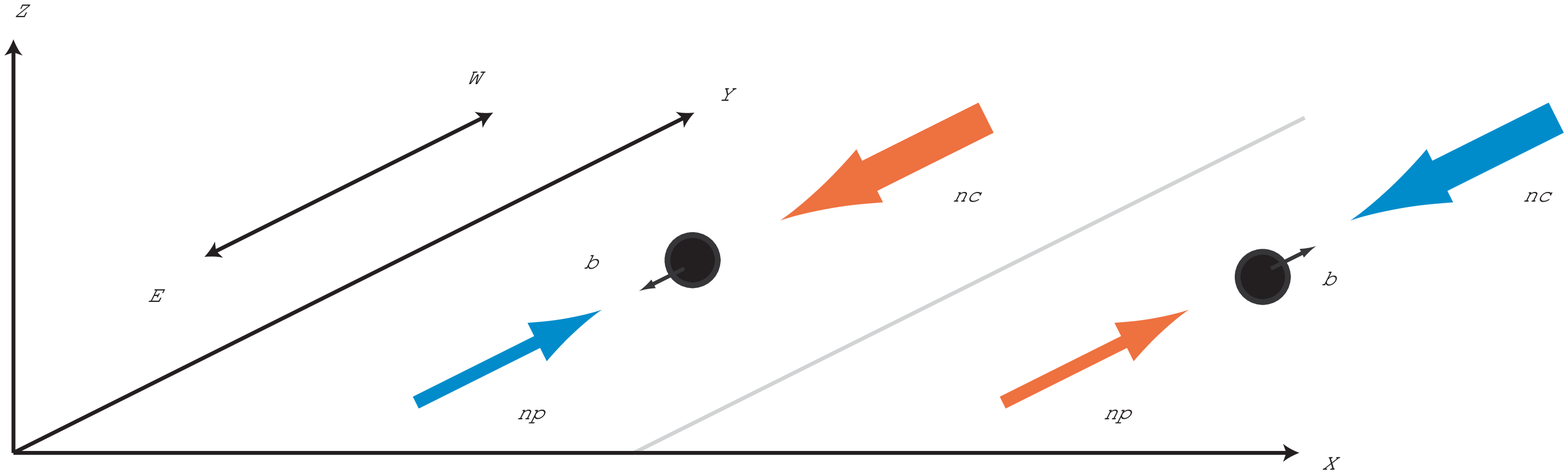}
\caption[The two systems generated in the cell]{The two systems generated in the cell. The experimental axis runs East$\leftrightarrow$West. Red and blue arrows are the red and blue detuned laser fields respectively. 
\label{fig:cellsetup}}
\end{center}
\end{figure}
This system can be modeled by two identical experiments with the probe and coupling field detunings interchanged between them, see Figure \ref{fig:cellsetup}. To begin the analysis we first consider a toy-model for this system which consists of two similar $\Lambda$-atoms, one propagating East and one propagating West, each with speed $\vv_{c}$. Assume that the $\Lambda$-atoms are identical except that the ground states states $\ket{1}$ and $\ket{2}$ switch roles between the true ground, and metastable ground states and that the appropriate transitions are addressed by the appropriate polarisations for each laser, see Figure \ref{fig:toymodel}. 
A finite value of $\kappa_{tr}$ manifests itself as an atomic-beam direction-dependent magnetic field, with the Lorentz violating contribution to the two-photon detunings given by:
\begin{align}
\label{eq:splitting1}
\Xi^{(E)} & =  -2 \nu_{0}\beta_{at} \kappa_{tr} \nonumber \\
\Xi^{(W)} & =  2 \nu_{0}\beta_{at} \kappa_{tr},
\end{align}
where the superscripts on $\Xi$ represent the propagation direction of the atomic beam.  A comparison between the line centres of two such systems could therefore be used to perform a differential measurement where there separation could be used to measure the size of $\kappa_{tr}$ according to,
\begin{equation} 
\label{eq:split1}
\Delta \Xi = |\Xi^{(E)} - \Xi^{(W)}| = 4 \kappa_{tr} \nu_{0}\beta_{at}.
\end{equation}
Looking for a separation in this way, rather than an absolute offset from zero, removes many of the systematic contributions to the uncertainty in the measurement.

In a real atomic system the direction of the shift, i.e. whether the line center moves to a positive of negative magnetic field, is also determined by the internal structure of each $\Lambda$-transition contributing to the system.

\section{Towards a real atomic system}
\begin{figure}[ht!]
\centering
\psfrag{A}{\Huge  A}
\psfrag{B}{\Huge  B}
\psfrag{DDa}{\scriptsize $\Delta_{1}$}
\psfrag{DDb}{\scriptsize $\Delta_{1}$}
\psfrag{$D_{1}$a}{\scriptsize $\Delta_{2}$}
\psfrag{$D_{1}$b}{\scriptsize $\Delta_{2}$}
\psfrag{n31}{\scriptsize $\nu_{31}$}
\psfrag{n32}{\scriptsize $\nu_{32}$}
\psfrag{n21}{\scriptsize $\nu_{21}$}
\psfrag{G31}{\scriptsize $\Gamma_{31}$}
\psfrag{G32}{\scriptsize $\Gamma_{32}$}
\psfrag{Opa}{\scriptsize $\Omega_{p}$}
\psfrag{Opb}{\scriptsize $\Omega_{p}$}
\psfrag{Oca}{\scriptsize $\Omega_{c}$}
\psfrag{Ocb}{\scriptsize $\Omega_{c}$}
\psfrag{1}{\scriptsize $\ket{1}$}
\psfrag{2}{\scriptsize $\ket{2}$}
\psfrag{3}{\scriptsize $\ket{3}$}
\psfrag{4}{\scriptsize $\ket{4}$}
\psfrag{5}{\scriptsize $\ket{5}$}
\psfrag{6}{\scriptsize $\ket{6}$}
\subfigure{\includegraphics[height=6.5cm]{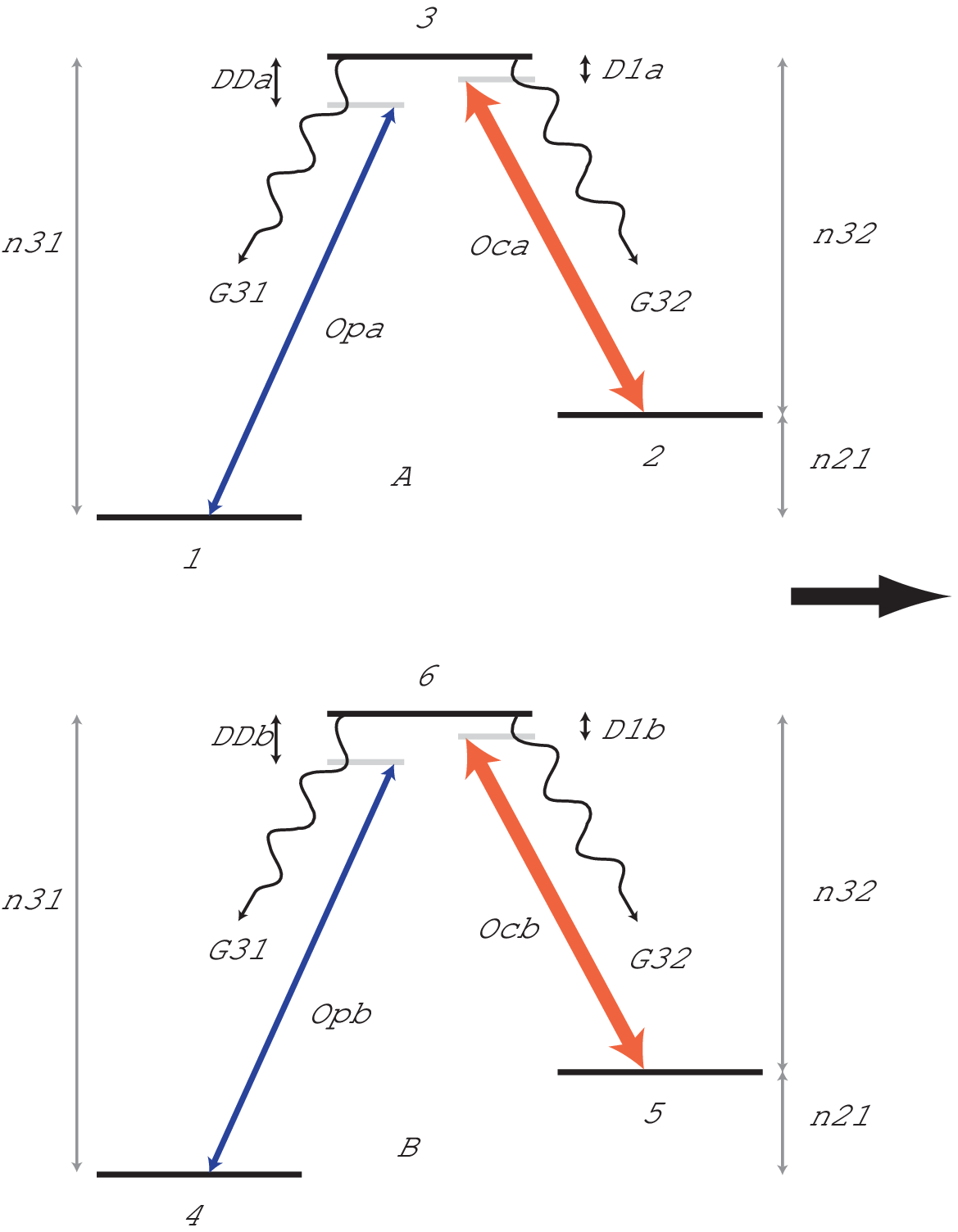}}
\psfrag{O1}{{\scriptsize{$\Omega_{c}^{(A)}$}}}
\psfrag{O2}{{\scriptsize{$\Omega_{c}^{(B)}$}}}
\psfrag{p1}{{\scriptsize{$\Omega_{p}^{(A)}$}}}
\psfrag{p2}{{\scriptsize{$\Omega_{p}^{(B)}$}}}
\psfrag{$D_{1}$}{{\scriptsize{$\Delta_{1}^{(A)}$}}}
\psfrag{d2}{{\scriptsize{$\Delta_{1}^{(A)}$}}}
\psfrag{$D_{1}$}{{\scriptsize{$\Delta_{2}^{(A)}$}}}
\psfrag{D2}{{\scriptsize{$\Delta_{2}^{(B)}$}}}
\psfrag{1}{\scriptsize $\ket{1}$}
\psfrag{2}{\scriptsize $\ket{2}$}
\psfrag{3}{\scriptsize $\ket{3}$}
\psfrag{4}{\scriptsize $\ket{4}$}
\psfrag{5}{\scriptsize $\ket{5}$}
\psfrag{6}{\scriptsize $\ket{6}$}
\subfigure{\includegraphics[height=6.5cm]{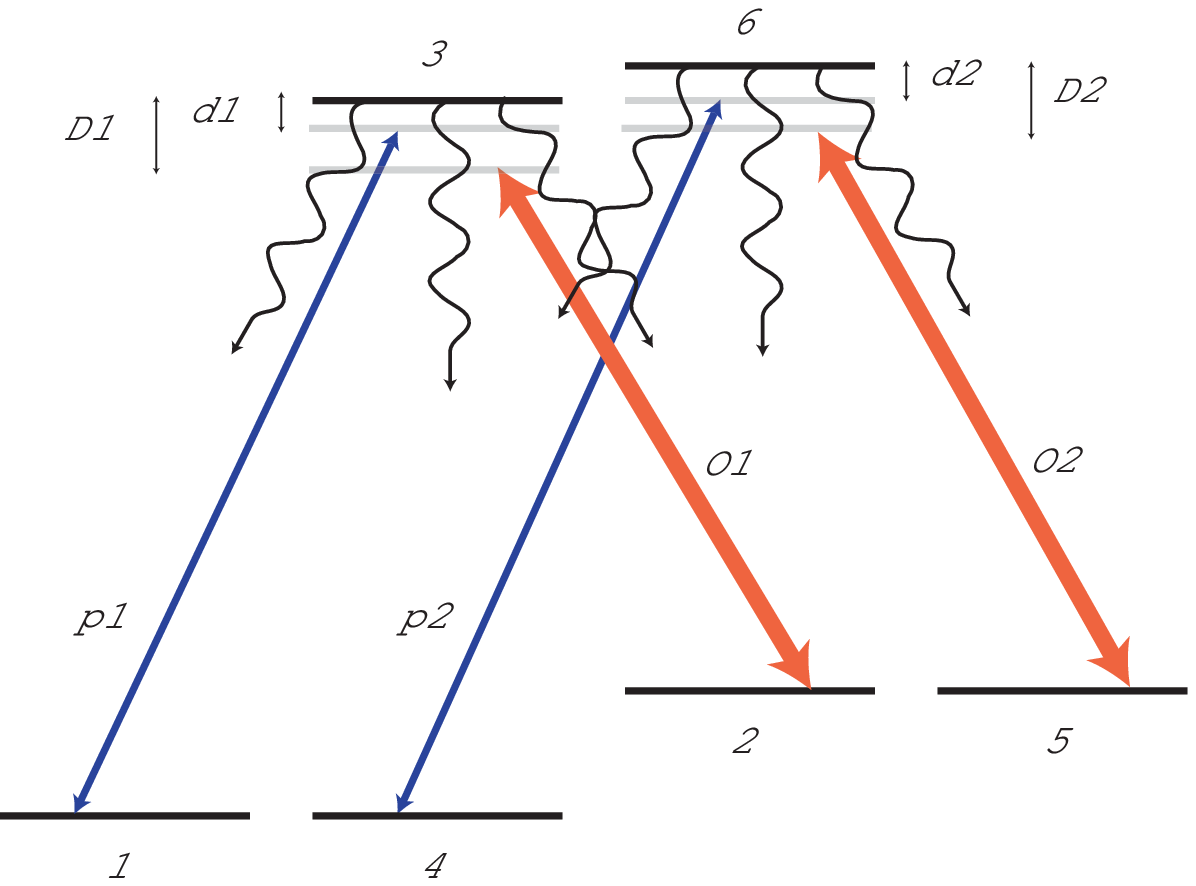}}
\caption[Two $\Lambda$-systems]{Two $\Lambda$-systems A and B formed between Zeeman sublevels $\ket{1}$, $\ket{2}$ and $\ket{3}$, and $\ket{4}$, $\ket{5}$ and $\ket{6}$ respectively. The states $\ket{1}$ and $\ket{4}$, $\ket{2}$ and $\ket{5}$, $\ket{3}$ and $\ket{6}$ are degenerate with one another. The excited state $\ket{3}$ can spontaneously decay (represented by curly arrows) to either one of the two ground states $\ket{1}$ or $\ket{4}$ at a rate of $\gamma_{31}$ or to the meta-stable state $\ket{2}$ at a rate of $\Gamma_{32}$, while the excited state $\ket{6}$ can decay to either one of the meta-stable states at a rate of $\Gamma_{32}$, or one the ground states $\ket{4}$ at a rate of $\Gamma_{31}$
\label{fig:2EIT}}
\end{figure}

So far we have described how EIT can be generated in an idealized atomic beam selected from a thermal distribution, however this is far from the situation encountered in a real atomic system. Before attempting to tackle the $D_{1}$ line in $^{85}$Rb (i.e. the system used in this experiment) we considere the somewhat simpler case of two $\Lambda$-systems formed between magnetic sublevels interacting with the same optical fields $\Omega_{p}$ and $\Omega_{c}$ as illustrated in  Figure \ref{fig:2EIT}.
In this system the energy levels $\ket{1}$ and $\ket{4}$, $\ket{2}$ and $\ket{5}$, $\ket{3}$ and $\ket{6}$ are degenerate with one another, except for some small detuning which can arise from an externally applied magnetic field. The Hamiltonian of the combined system can therefore be written as,
\begin{align}
\widetilde{\mathscr{H}}_{2} &= \left(  \begin{array}{cc} \mathscr{H}_{A} & 0 \\ 0 & \mathscr{H}_{B}  \end{array} \right),
\end{align}
where the diagonal terms are $3\times3$ matrices with the appropriate choice of detunings and Rabi fields for the respective $\Lambda$-system inserted. To examine this system in more detail we construct a Master equation using all six energy levels making sure to also include an appropriately modified decay operator,
\begin{align} 
D{\rho} & = \frac{1}{4}[2\Gamma_{31}\rho_{33}\ket{1}\bra{1} + 2\Gamma_{32}\rho_{33}\ket{2}\bra{2} - 2 (\Gamma_{31} + \Gamma_{32})\rho_{33}\ket{3}\bra{3}\nonumber \\
&+ 2\Gamma_{31}\rho_{66}\ket{4}\bra{4} + 2\Gamma_{32}\rho_{66}\ket{5}\bra{5} - 2 (\Gamma_{31} + \Gamma_{32})\rho_{66}\ket{6}\bra{6} \nonumber \\
&- \gamma_{21}(\rho_{21}\ket{2}\bra{1} + \rho_{51}\ket{5}\bra{1} + \rho_{42}\ket{4}\bra{1}) \nonumber \\
&- \gamma_{31}(\rho_{31}\ket{3}\bra{1} + \rho_{61}\ket{6}\bra{1} + \rho_{42}\ket{4}\bra{3}) \nonumber \\
&- \gamma_{32}(\rho_{32}\ket{3}\bra{2} + \rho_{62}\ket{6}\bra{2} + \rho_{42}\ket{5}\bra{3}) \nonumber \\
&- \gamma_{41}\rho_{41}\ket{4}\bra{1} - \gamma_{52}\rho_{52}\ket{5}\bra{2}- \gamma_{63}\rho_{63}\ket{6}\bra{3}] + H.c.
\end{align}
where we have assumed that the two ground states $\ket{1}$ and $\ket{4}$ to remain equally populated, $\rho_{11}\simeq\rho_{44}\simeq1/2$; a reasonable assumption for such a system inside an atomic vapour cell where the population is thermally distributed between the two ground states. Therefore the general density matrix for this system can be simplified slightly to become, 
\begin{align}
\label{eq:6way_denmat}
\rho &= \begin{pmatrix} 
\begin{pmatrix} 1/2 & \rho_{12} & \rho_{13} \\ \rho_{21} & 0 &\rho_{23}  \\ \rho_{31} & \rho_{32} &0 \end{pmatrix} 
& \begin{pmatrix} \rho_{14} & \rho_{15} & \rho_{16} \\ \rho_{24} & \rho_{25} & \rho_{26} \\ \rho_{34} & \rho_{35} & \rho_{36} \\ \end{pmatrix} \\
\begin{pmatrix} \rho_{41} &\rho_{42} & \rho_{43} \\  \rho_{51} & \rho_{52} & \rho_{53} \\ \rho_{61} & \rho_{62} & \rho_{63} \end{pmatrix} 
& \begin{pmatrix} 1/2 & \rho_{45} & \rho_{46} \\ \rho_{54} & 0 & \rho_{56} \\ \rho_{64} & \rho_{65} & 0\\ \end{pmatrix}
\end{pmatrix}.
\end{align}
Solving the Master equation for this density operator we find that the time evolution of the coherences occurring in the block wise diagonal terms depend only upon the population of their respective quadrant. That is to say that $\rho_{31} = f(\rho_{11}) \ne f(\rho_{44})$, $\rho_{64} = f(\rho_{44}) \ne f(\rho_{11}) $ etc. Additionally, because $\ket{1}$ and $\ket{4}$ are connected to $\ket{3}$, and $\ket{2}$ and $\ket{5}$ are connected to $\ket{6}$ via different spontaneous emission polarisation channels no coherence can be formed between them under normal operation. Because the time evolution of the blockwise, off diagonal terms, in equation (\ref{eq:6way_denmat}) does not depend on the level populations, if there is no coherence in the first instance, then there can be not be coherence at any time there after. This permits further simplification of the density operator,
\begin{align}
\label{eq:red_den_big}
\rho &= \begin{pmatrix} 
\begin{pmatrix} 1/2 & \rho_{12} & \rho_{13} \\ \rho_{21} & 0 &\rho_{23}  \\ \rho_{31} & \rho_{32} &0 \end{pmatrix} 
& \begin{pmatrix} 0 & 0 & 0 \\ 0 & 0 & 0 \\ 0 & 0 & 0 \\ \end{pmatrix} \\ \begin{pmatrix}0 &0 & 0 \\  0 & 0 & 0 \\ 0 & 0 & 0 \end{pmatrix} 
& \begin{pmatrix} 1/2 & \rho_{45} & \rho_{46} \\ \rho_{54} & 0 & \rho_{56} \\ \rho_{64} & \rho_{65} & 0\\ \end{pmatrix}
\end{pmatrix}.
\end{align}
Because the elements of the blockwise diagonal quadrants in equation \ref{eq:red_den_big} do not depend on any terms contained within the other diagonal quadrant, or the off-diagonal quadrants the two systems A and B can be considered independent of one another. Therefore a general solution for the steady state density matrix of the combined system is, 
\begin{equation}
\rho_{SS} = P_{A} \rho_{SS}^{(A)} + P_{B} \rho_{SS}^{(B)},
\end{equation}
where $P_{A}$ and $P_{B}$ are the probabilities of being in systems A and B respectively and $\rho_{SS}^{(A)}$ and $\rho_{SS}^{(B)}$ are the steady state density matrices for the two systems if they were isolated. Assuming equal probabilities $P_{A}$ and $P_{B}$ the effective susceptibility of the two systems is,
\begin{align}
\chi^{(1)} &= \dfrac{2N}{\epsilon_{0}E_{0}V}\left( \mu_{13}\rho_{31} + \mu_{46}\rho_{64} +  \mu_{23}\rho_{32} + \mu_{56}\rho_{65}\right) \\ \nonumber
~ &= \chi^{(1)}(A) + \chi^{(1)}(B),
\end{align}
and therefore the effective susceptibility of this combined system is simply that of the sum of the two systems A and B. In the $D_{1}$ line of $^{85}$Rb there are four possible $\Lambda$-systems which can be formed using circularly polarised light. We assume that these four systems can be treated independently and that the total susceptibility observed will be the sum of the susceptibilities of the independent systems.
\begin{equation}
\chi^{(1)}_{total} = \displaystyle \sum_{j} \chi^{(1)}_{j}
\end{equation}

\subsection{EIT in the $D_{1}$ line of $^{85}$Rb vapour}
The Maxwell-Boltzmann velocity vector distribution is symmetrical in velocity, therefore an atom travelling along the Easterly unit vector in the laboratory frame with a velocity $\vv_{c}$ is as equally likely as an atom travelling along a Westerly unit vector at the same speed, for example. Because the rest frame laser frequency has been chosen such that a symmetrical Doppler shift brings the two lasers into resonance with their respective transition there will be two effective beams which can be selected from the thermal distribution.
\begin{figure}[ht!]
\centering
\subfigure[Thermal population]{\includegraphics[width = 5cm]{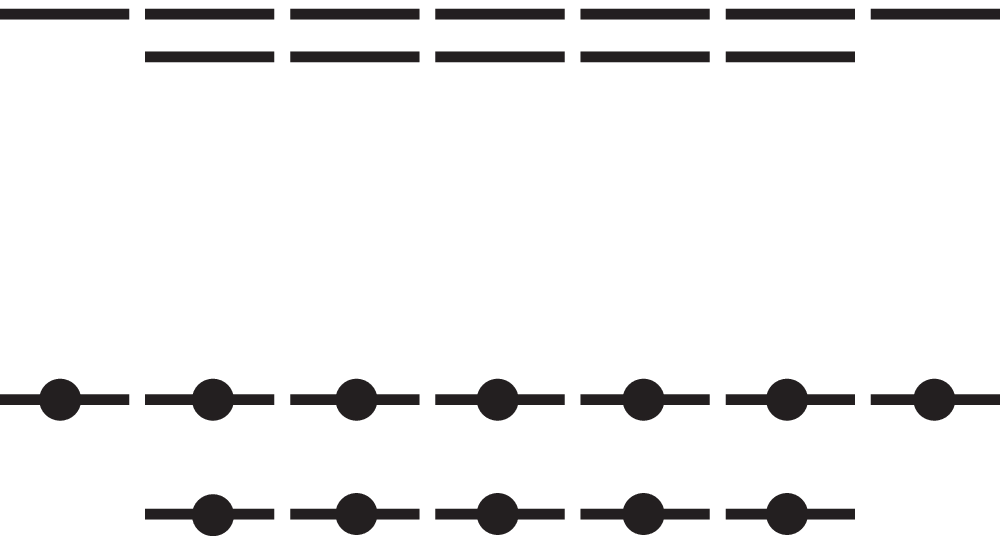}}
\psfrag{F2}{\tiny $F = 2$}
\psfrag{F3}{\tiny $F = 3$}
\psfrag{m1}{\tiny $\ket{\Omega_{p},\sigma_{-}}$}
\psfrag{m2}{~}
\psfrag{m3}{~}
\psfrag{m4}{\tiny $\ket{\Omega_{c},\sigma_{+}}$}
\psfrag{v}{\tiny $\vv_{at}$}
\subfigure[Contributing population]{\includegraphics[width = 12cm]{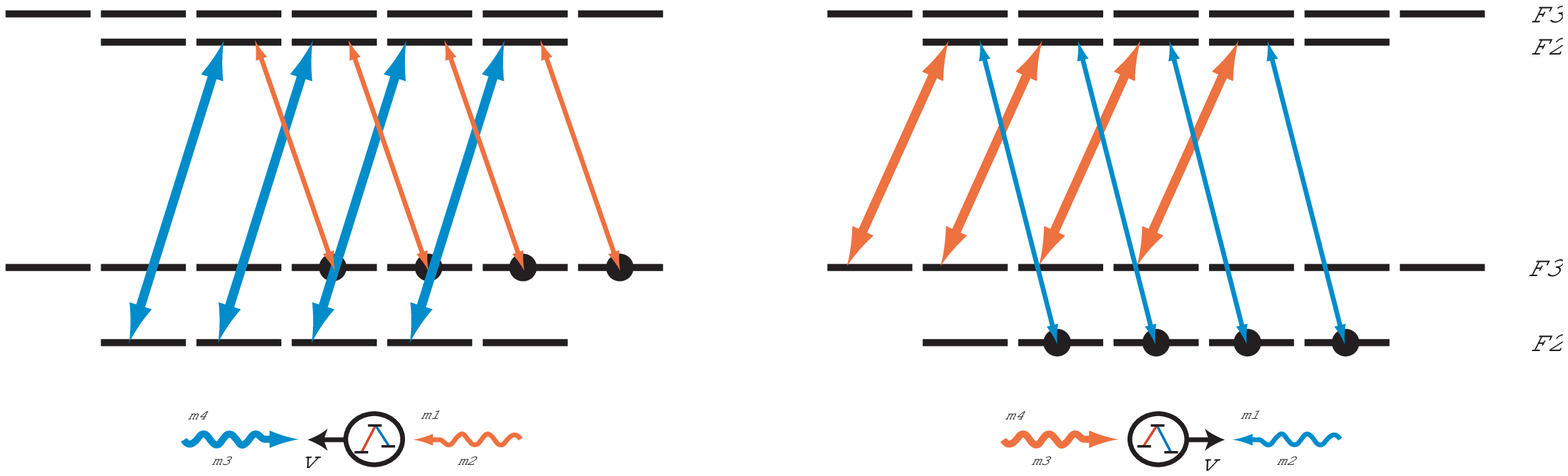}}
\caption[The population in the cell]{The population in the cell. is initially thermally populated between the $F = 2$ and $F = 3$ ground states of the $5^{2}S_{1/2}$ manifold. However, only those atoms with their valance electron in an appropriate state will contribute to any observed EIT signal with any other atoms effecting only the background absorption which changes only on a frequency scale much larger than that of EIT.
\label{fig:population1}}
\end{figure}
The population inside the cell is initially thermally distributed, this means there are equal populations of atoms with their valance electrons in each one of the possible $m_{F}$ magnetic sublevels within the $F = 2$ and $F = 3$ manifolds of the ground state. The description previously given to describe the susceptibility of an idealized three level system assumed that the population was initially all in the absolute ground state. Although this is not really the case, only those atoms with their valance electron in an appropriate hyperfine state for their orientation with respect to the laser fields will contribute to the observed EIT signal. All other atoms can only contribute to the background absorption, which changes only on a frequency scale much larger than that of EIT. Any optical pumping occurring will only increase the population's capable of exhibiting EIT. 

The separation between the $F = 3$ and $F = 2$ states in the $5^{2}S_{1/2}$ manifold is $\sim 360$MHz which is large enough to individually probe either of these states without influence from the other. Considering only the $F = 2$ excited state there should be $8$ different $\Lambda$-systems which can contribute to any observed EIT spectrum; $4$ in each of the two atomic beams propagating along the experimental axis, see Figure \ref{fig:population1}.  Although each $\Lambda$-system interacts with the same optical fields, due to different dipole matrix elements as well as different gryomagnetic ratios and magnetic quantum numbers each system can support a different linewidth. Regardless of the contributing $\Lambda$-system, the line center of each resultant spectrum should be centred at zero magnetic field.

\subsection{Simulating a violation}
The experiment uses two beams of $^{85}$Rb atoms propagating in opposite directions along the East$\leftrightarrow$West experimental axis. The predicted EIT lineshapes for all eight of the $\Lambda$-systems which can be supported by the $\ket{1} = \ket{F = 2}$, $\ket{2} = \ket{F = 3}$ and $\ket{3} = \ket{F = 2}$ states within the $D1$ manifold of $^{85}$Rb have been modelled. 
\begin{figure}[ht!]
\centering
\psfrag{F2}{\scriptsize $F = 2$}
\psfrag{F3}{\scriptsize $F = 3$}
\psfrag{m1}{\scriptsize $\ket{\Omega_{p},\sigma_{-}}$}
\psfrag{m2}{~}
\psfrag{m3}{~}
\psfrag{m4}{\scriptsize $\ket{\Omega_{c},\sigma_{+}}$}
\psfrag{v}{\scriptsize $\vv_{at}$}
\psfrag{1}{\textcircled{\small $1$}}
\psfrag{2}{\textcircled{\small $2$}}
\psfrag{3}{\textcircled{\small $3$}}
\psfrag{4}{\textcircled{\small $4$}}
\psfrag{E}{East}
\psfrag{W}{West}
\includegraphics[width = 14cm]{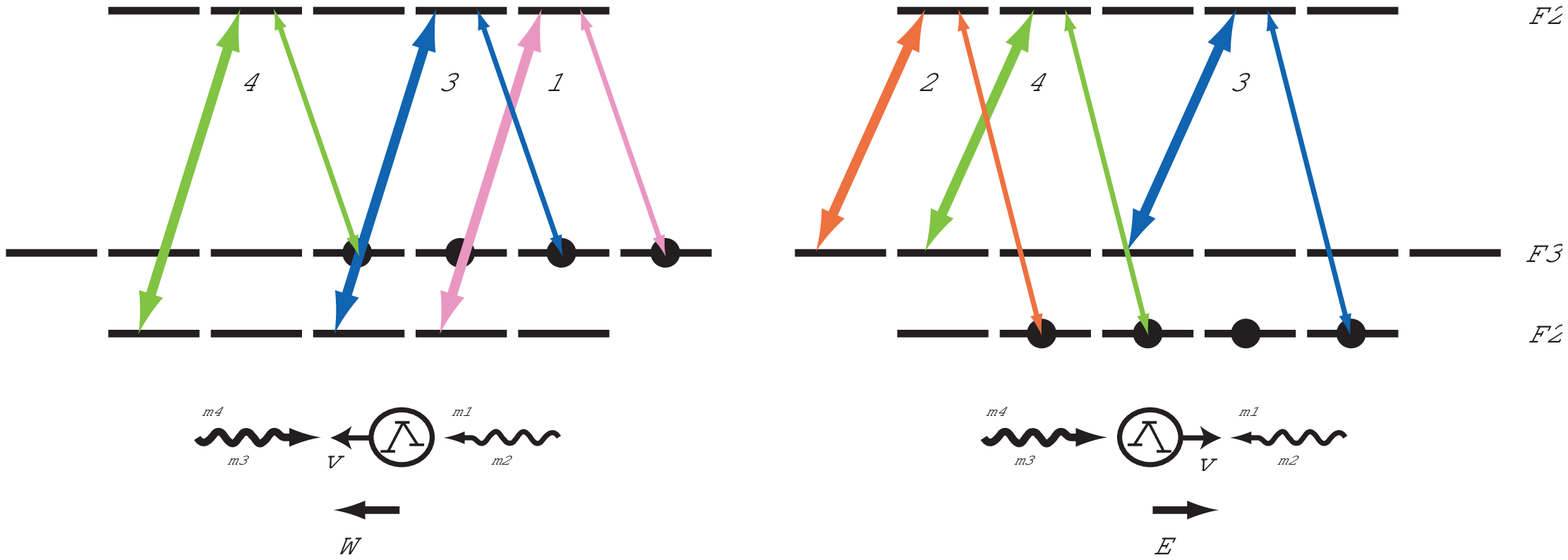}
\caption[Six $\Lambda$-systems in two atomic beams]{The six $\Lambda$-systems which can be generated in the two atomic beams inside the cell. The colour coding of the $\Lambda$-systems corresponds to the spectra depicted in Figure \ref{fig:EIT_dip1}. The colour coding corresponds to the predicted spectra that can be found in Figure \ref{fig:EIT_dip1}
\label{fig:many_eit2}}
\end{figure} 
Of these eight $\Lambda$-systems only six can contribute to any EIT signal, because the internal structure is such that there is no magnetic field which can generate a detuning for those systems with $\ket{3} = \ket{m_{F}=0}$. Of the remaining six contributing systems, see Figure \ref{fig:many_eit2}, only four provide unique deviations of line centre from zero magnetic field: For finite values of $\kappa_{tr}$ transitions with $\ket{3} = \ket{m_{F} = 1}$, are equivalent to one another in counter-propagating atomic beams. This also applies to transitions with $\ket{3} = \ket{m_{F} = -1}$.

\begin{figure}[ht!]
\centering
%
%
\begin{psfrags}%
\psfragscanon%
%
\psfrag{s05}[t][t]{\color[rgb]{0,0,0}\setlength{\tabcolsep}{0pt}\begin{tabular}{c}$B/nT$\end{tabular}}%
\psfrag{s06}[b][b]{\color[rgb]{0,0,0}\setlength{\tabcolsep}{0pt}\begin{tabular}{c} Absorption/arb. un\end{tabular}}%
\psfrag{s10}[][]{\color[rgb]{0,0,0}\setlength{\tabcolsep}{0pt}\begin{tabular}{c} \end{tabular}}%
\psfrag{s11}[][]{\color[rgb]{0,0,0}\setlength{\tabcolsep}{0pt}\begin{tabular}{c} \end{tabular}}%
\psfrag{s12}[l][l]{\color[rgb]{0,0,0}{\scriptsize \textcircled{{$4$}}}}%
\psfrag{s13}[l][l]{\color[rgb]{0,0,0}{\scriptsize \textcircled{{$1$}}}}%
\psfrag{s14}[l][l]{\color[rgb]{0,0,0}{\scriptsize \textcircled{{$2$}}}}%
\psfrag{s15}[l][l]{\color[rgb]{0,0,0}{\scriptsize \textcircled{{$3$}}}}%
\psfrag{s16}[l][l]{\color[rgb]{0,0,0}{\scriptsize \textcircled{{$4$}}}}%
\psfrag{x01}[t][t]{$-60$}%
\psfrag{x02}[t][t]{$-40$}%
\psfrag{x03}[t][t]{$-20$}%
\psfrag{x04}[t][t]{$0$}%
\psfrag{x05}[t][t]{$20$}%
\psfrag{x06}[t][t]{$40$}%
\psfrag{x07}[t][t]{$60$}%
%
\psfrag{v01}[r][r]{~}
\psfrag{v02}[r][r]{~}
\psfrag{v03}[r][r]{~}
\psfrag{v04}[r][r]{~}
\psfrag{v05}[r][r]{~}
\psfrag{v06}[r][r]{~}
\psfrag{v07}[r][r]{~}
\psfrag{v08}[r][r]{~}
%
\includegraphics[width=13cm]{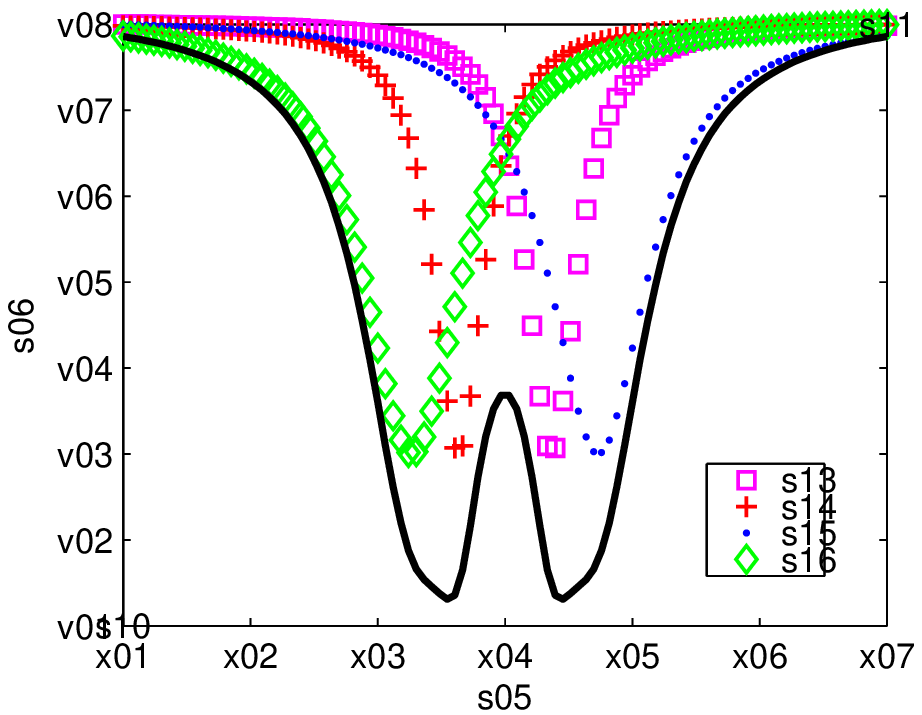}%
\end{psfrags}%
%

\caption[Simulating a violation]{Simulating a violation. Displayed are the four contributing  spectra. The atomic parameters used in these simulations were matched to those obtained by fitting to current experimental data, where we have assumed a Lorentz violation of $\kappa_{tr}=8\times 10^{-8}$ (the current best experimental limit). The $({\blue \cdot~\cdot~\cdot~\cdot})$ and $({\green \Diamond \Diamond \Diamond \Diamond})$ represent transitions between $\ket{3}=\ket{m_{F}=\pm1}$ in both velocity classes. Because of the symmetrical structure of these transitions there is an additional factor of two contribution from these states. The $({\magenta \Box \Box \Box \Box})$ and ({\red $+$ $+$ $+$ $+$}) represent the $\ket{3}=\ket{m_{F}=2}$ transition in the counter propagating atomic beams. The solid black line is the summed contribution from all relevant transitions and is the expected output from the experiment assuming an equal contribution from all $\Lambda$-configurations.
\label{fig:EIT_dip1}}
\end{figure}

The four remaining unique spectra which occur for $\kappa_{tr} \ne 0$ are shown in Figure \ref{fig:EIT_dip1}.
The linewidths of the induvidual spectra in Figure \ref{fig:EIT_dip1} differ due to their varying Clebsch-Gordon (CG) coefficients, magnetic quantum numbers and gyromagnetic ratios. The $\Lambda$-systems \textcircled{\small $1$} and \textcircled{\small $2$} are twice as sensitive to magnetic fields as \textcircled{\small $3$} and \textcircled{\small $4$}. The simulation assumed a value of $\kappa_{tr} = 8 \times 10^{-8}$ and a fixed intensity coupling field from which the appropriate CG coefficient were derived. All frequency parameters used in this simulation correspond to the $D1$ line of $^{85}$Rb.

The results of this simulation are presented in Figure \ref{fig:EIT_dip1}. This figure shows each of the four unique spectra, their colours are matches to the appropriate $\Lambda$-systems shown in Figure \ref{fig:many_eit2}. It also shows the summation of all these spectra (black line) taking into account the factor of two contribution in the amplitude of the most magnetically insensitive transitions \textcircled{\small $3$} and \textcircled{\small $4$}. What this figure shows is that for large enough values of $\kappa_{tr}$ the resultant spectrum from the four contributing unique spectra also exhibits a splitting. It also shows that for this particular system the separation between resultant peaks in the overall spectrum is governed by the splitting between the most magnetically sensitive $\Lambda$-systems \textcircled{\small $1$} and \textcircled{\small $2$}.

Despite the added complexity of this system over the toy-model presented in Figure \ref{fig:toymodel} it is still possible to access $\kappa_{tr}$ according to equation (\ref{eq:split1}) where the splitting $\Delta \Xi$ is now determined by the splitting of the most magnetically sensitive $\Lambda$-systems \textcircled{\small $1$} and \textcircled{\small $2$}.

The dominant contribution to any Lorentz violating signal observed in this experiment will originate from the SME parameter $\kappa_{tr}$.  Although there will be additional contributions from the other SME parameter, $\kappa_{o+}$, $\kappa_{e-}$, $\kappa_{o-}$ and $\kappa_{e+}$ which have so far been neglected because of the significantly tighter bounds placed on them by other experiments. 

In this paper we have presented a detailed examination of the Ives-Stilwell experiment in the SME for a real atomic system. It shows that the Modified Ives-Stilwell experiment can provide a sensitive test of Lorentz Invariance. We have shown that for a relatively simple system it is possible to exceed the best current experimental test for $\kappa_tr$. More advanced setups would allow considerably better measurements, for example using counter propagating atomic beams, rather than a gas cell would enable a further 100 fold increase in sensitvity.

\section{Aknowledgements}
We wish to aknowledge the support of the EPSRC in the form of an  Advanced Fellowship for BV and a DTC award for JC, and the STFC for providing startup support via PPRP grant 327.

\addcontentsline{toc}{chapter}{Bibliography}
\bibliographystyle{apsrev}
\bibliography{Bibliography}

\end{document}